# Massively parallel optical-soliton reactors

W. He[1,*], M. Pang[1], D. H. Yeh[1], J. Huang[1] and P. St.J. Russell[1,2]

[1]Max Planck Institute for the Science of Light and [2]Department of Physics, Friedrich-Alexander-Universität, Staudtstrasse 2, 91058 Erlangen, Germany

[*]Corresponding author: wenbin.he@mpl.mpg.de

**Mode-locked lasers have been widely used to explore interactions between optical solitons, including bound-soliton states that may be regarded as "photonic molecules". Conventional mode-locked lasers can however host at most only a few solitons, which means that stochastic behaviour involving large numbers of solitons cannot easily be studied under controlled experimental conditions. Here we report the use of an optoacoustically mode-locked fibre laser to create hundreds of temporal traps or "reactors" within which multiple solitons can be isolated and controlled both globally and individually. We achieve on-demand synthesis and dissociation of soliton molecules within these parallel reactors, in this way unfolding a novel panorama of diverse dynamics in which the statistics of multi-soliton interactions can be studied. The results are of potential importance in all-optical information processing and in the control of ultrafast pulsed lasers. (135 words)**

The cavity of a mode-locked laser, in which particle-like optical solitons can reside, has been widely-used as a platform for investigating complex soliton dynamics[1–8]. Recent studies using time-stretched dispersive Fourier transformation (TS-DFT)[9] have suggested that a useful analogy can be drawn between multi-soliton complexes bound by soliton-soliton interactions and molecules[3,4,8,10–13]. Such "soliton molecules[14–16]" display complex internal dynamics[3,4,10,17–19] and have attracted considerable interest both in fundamental nonlinear science and in applications such as ultrafast lasers[1,20,21], optical communications[22,23] and all-optical information processing[10,24].

In previous work, soliton molecules have mostly been investigated as single entities generated by random excitations in optical cavities[17,18], whereas real molecules participate in huge numbers in dynamic processes such as chemical reactions. Conventional mode-locked lasers are generally able to host only few solitons[1,20], so that only rarely have they been used to study the statistics of nonlinear soliton dynamics[2,17,25–28]. To be successful, such studies demand control of the interactions between large numbers of solitons and soliton complexes, something that is experimentally challenging[8,10,24]. In particular, the synthesis in mode-locked lasers of soliton-molecules from single solitons, and their dissociation into single solitons[1] (which can be regarded as analogous with chemical reactions between atoms) has not yet been realized in a controllable manner.

The physical scale of a mode-locked fibre laser is typically many times longer than an individual soliton, permitting in principle the coexistence of very large numbers of solitons and soliton molecules. In practice, however, uncontrollable drifting and collisions caused by noise[27,29–31], together with intrinsic group velocity differences between solitonic elements[32–35], has greatly limited the flexibility of mode-locked lasers as hosts for large solitonic structures. We have previously reported that the optomechanical lattice created in an optoacoustically mode-locked fibre laser[24,36] by a short length of photonic crystal fibre (PCF)[37] can synchronize the velocities of intra-cavity solitonic elements through long-range soliton interactions. Several solitonic elements can be temporally isolated and stably trapped within each time-slot of the optomechanical lattice, forming a stable supramolecular structure[8]. Consecutive time-slots can thus function as parallel soliton reactors, echoing experiments on controlling chemical reactions by atomic manipulation[38]. By deliberately initiating the formation and dissociation of soliton molecules in these parallel reactors globally and individually, we are able to uncover previously unexplored stochastic aspects of soliton-soliton interactions. The build-up of a stable soliton molecule, which we observe in the experiments, generally requires multiple collisions[39–41], while the break-up of soliton molecules occurs in many difference ways. Statistical analysis also reveals that the reaction rates of both processes follow classical chemical kinetics. All-optical control enables reversible and selective switching between different multi-soliton states in each reactor, suggesting applications in information storage, data processing and logical operations using solitons as bits[10,23,24,42].

**Basic concept of parallel optical-soliton reactors**

The optoacoustically mode-locked fibre laser built to host and control the parallel optical-soliton reactors is briefly sketched in Fig.1a (see details in Supplementary Note 1). A 2-m-long solid-core PCF (core-structure in Fig.1b) supporting an $R_{01}$-like mechanical resonance[37] at 1.887 GHz is inserted into an Er-doped fibre laser cavity (Fig.1c). The optically-driven acoustic vibration in the PCF core divides the ~20 m cavity (~104 ns round-trip time) into 195 time-slots ~532 ps wide (one acoustic cycle), creating an optomechanical lattice[24]. At pump powers above 600 mW, this lattice can be adjusted to host a variety of solitonic supramolecules[8]. Each time-slot of the optomechanical lattice, functioning as a potential well, can stably trap one or more solitons bound by long-range and/or short-range forces (see Fig.1d), while the entire structure can accommodate a large population of solitons and soliton molecules, all sharing the same group velocity.

Long-range bound, phase-uncorrelated soliton states[8] are ideal start- and end-points for the synthesis and dissociation of phase-locked soliton molecules. In order to initiate soliton reactions, long-range repulsive forces between solitons must be controlled either globally or individually. As shown in Fig.1a, global control is achieved mainly by perturbing specific laser parameters, such as gain, loss, and cavity length (see Supplementary Note 1). As shown in Fig.1d, abrupt attenuation of long-range repulsive forces within the traps can cause the solitons to move towards each other, resulting in multiple collisions and eventually the formation of a soliton molecule. Conversely, soliton molecules can be broken up by enhancing the repulsive forces, causing the solitons to dissociate and settle down into a phase-uncorrelated long-range bound state. Reactions can be triggered in individual traps by launching addressing pulses into the laser cavity to perturb the soliton interactions by cross-phase modulation (XPM)[24] (see Fig.1a). Individual control allows each solitonic element to be reversibly and selectively edited (see results below). Thus the optoacoustic traps can host controllable reactions between solitonic elements in a manner reminiscent of chemical reactors (see Fig.1e).

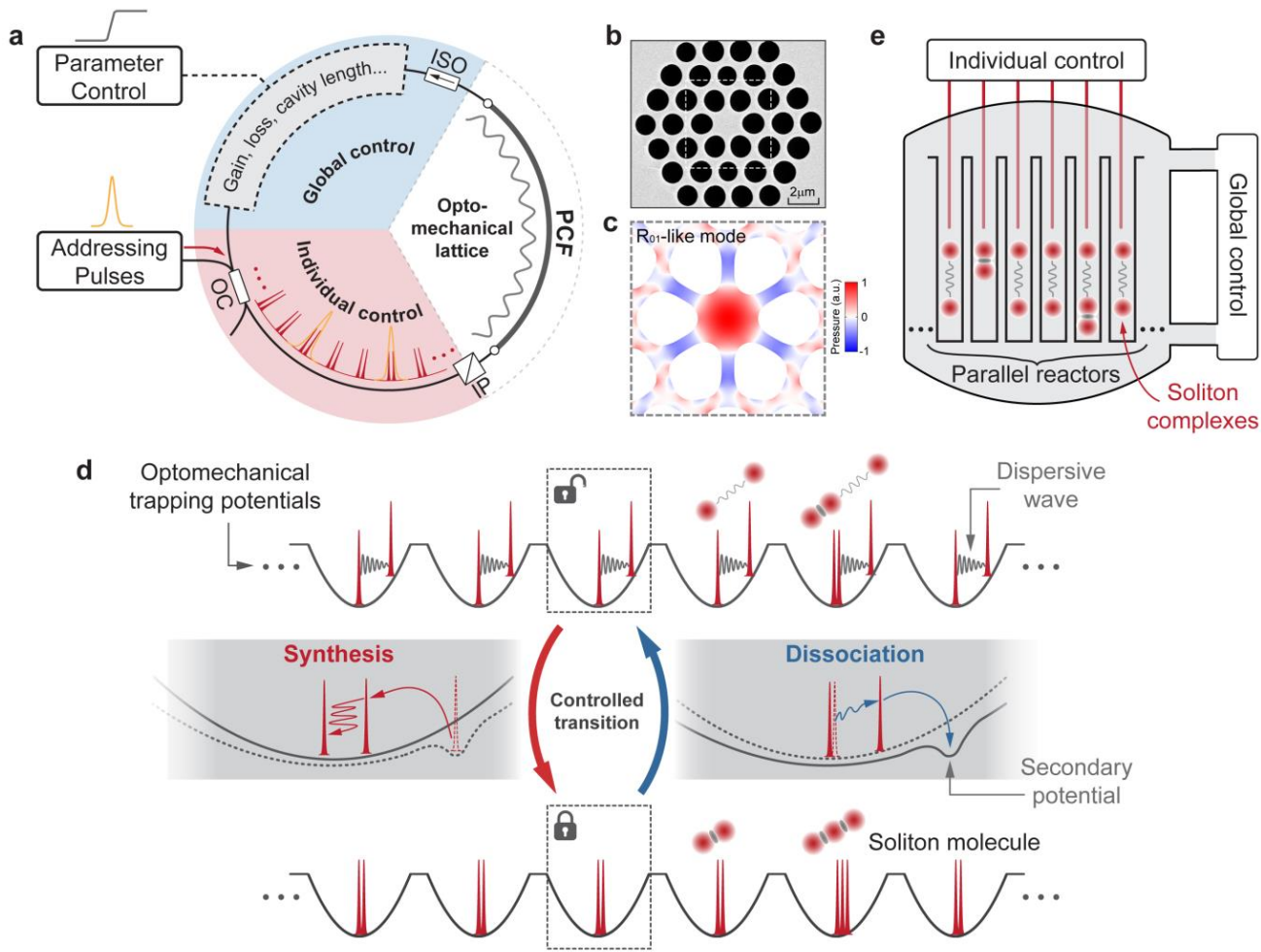

**Fig.1 | Conceptual illustration of parallel optical-soliton reactors in a fibre laser cavity. a** Schematic of the mode-locked fibre laser cavity. IP: inline polariser, OC: optical coupler, ISO: isolator. The acoustic resonance in the PCF core creates an optomechanical lattice that divides the laser cavity into many separate time-slots, each hosting a few solitons. In order to initiate soliton reactions in these time-slots, the laser parameters can be perturbed globally or by addressing individual time-slots with carefully timed external pulses. **b** Scanning electron micrograph (SEM) of the PCF microstructure. **c** Finite-element simulation of the $R_{01}$-like acoustic mode in the PCF core. The displacement is exaggerated for clarity and the normalized pressure is indicated by the colour map. **d** Schematic of controlled soliton reactions in the optomechanical lattice. The solitonic elements trapped in each reactor can be transitioned between phase-uncorrelated long-range bound states and phase-locked soliton molecules, corresponding to the synthesis and dissociation of soliton molecules. **e** Sketch of a system of parallel soliton reactors with global and individual control, mimicking a chemical reactor.

## Formation of soliton molecules via multiple collisions

To investigate the dynamics of soliton-molecule formation, we prepared a stable soliton supramolecule (see Methods) as the initial state (Fig.2a), consisting of 195 time-slots, all with two long-range-bound solitons, except for a reference slot containing a single soliton (see Fig.2b). In the absence of repulsive forces, the solitons within a single time-slot will tend to collide. This is prevented in practice by repulsive forces that arise from the shedding of dispersive waves[30,43–45], leading to formation of a secondary trapping potential (Fig.1d) that causes long-range binding at ~60 ps separation (Figs. 2b and 2c)—much longer that the ~1 ps duration of the solitons (see Supplementary Note 3). As a result the two trapped solitons have negligible field overlap and are thus uncorrelated in phase[8,30].

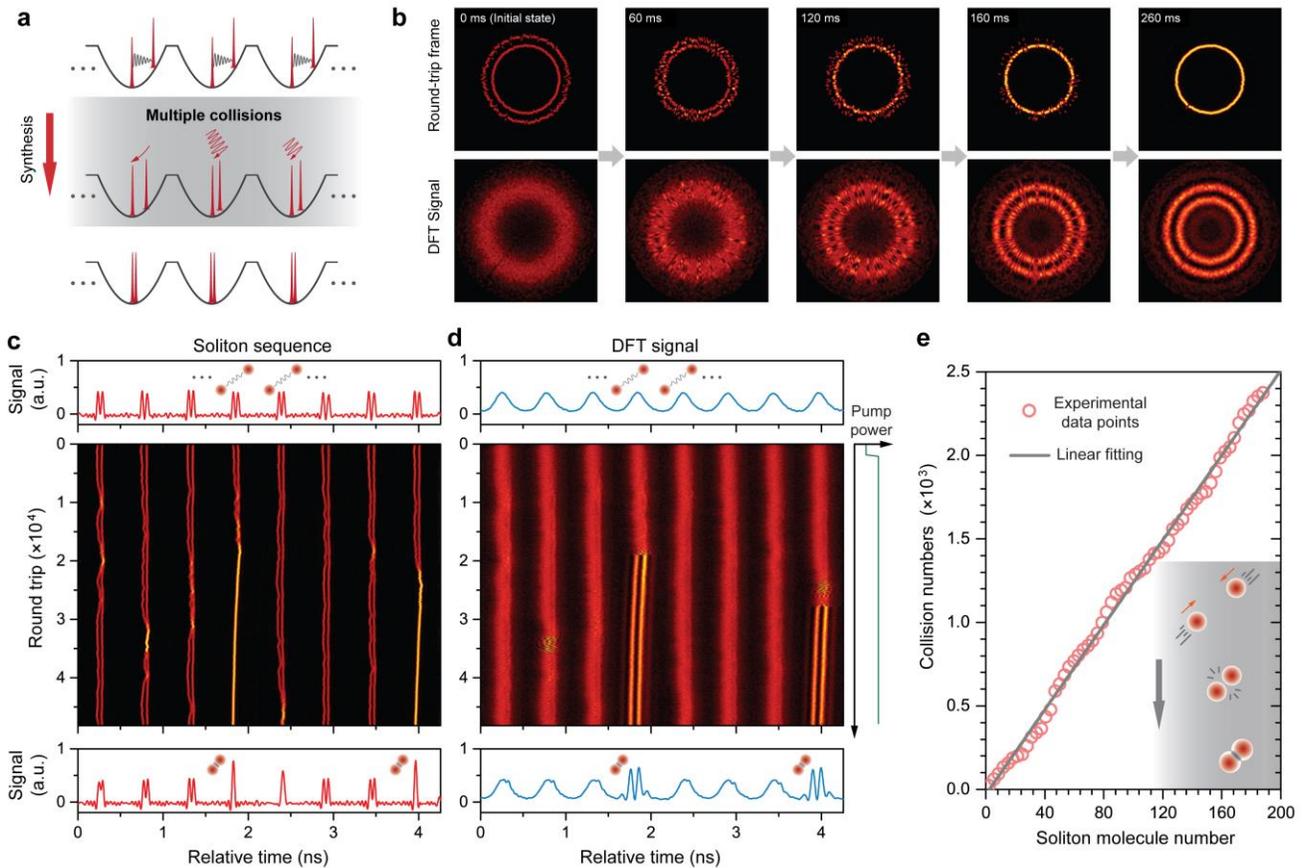

**Fig.2 | Global synthesis of phase-locked soliton molecules. a** Schematic of the synthesis of soliton molecules in parallel reactors. Two solitons are initially prepared in a long-range phase-uncorrelated bound state. After an abrupt change in pump power or cavity loss, the long-range binding collapses and the two solitons experience multiple collisions before forming a stable soliton molecule. **b** Upper row: Selected frames from an experimental recording of the synthesis process over all 195 reactors, plotted in cylindrical coordinates (see Methods and Supplementary Note 2). Lower: The corresponding DFT signal. The initial state contains 194 time-slots with two long-range bound solitons and a reference time-slot with a single soliton. The gradual establishment of spectral fringes in the DFT signal in all time-slots indicates formation of soliton molecules. See Supplementary Movie 1 for the complete recording. **c** Time-domain evolution in 8 consecutive time-slots over the initial 49,000 round-trips (~5 ms). **d** The corresponding DFT signal. Soliton molecules have formed in only two time-slots, as indicated by the sharp fringes in **d**. Top and bottom panels in **c** and **d** show the recorded signals for the initial and final round-trips. **e** Cumulative number of soliton collisions is proportional to the number of soliton molecules in all 195 reactors during a single synthesis (red circles). The grey line is a linear fit.

Synthesis is initiated by abruptly changing the pump power or the intra-cavity loss, which causes a decrease in the inter-soliton repulsive force (see Supplementary Note 3 and 4) and a gradual reduction in soliton spacing in all the reactors, culminating in multiple soliton collisions (Fig.2a). We studied the real-time motion of the reacting solitons by time-domain recording (limited by the bandwidth of the photodiode, so suitable only for long-range binding) and DFT (suitable for short-range <15 ps binding) in all the reactors (see Methods). The entire reaction process over all time-slots, initiated by pump perturbation, was first recorded at 5 kHz frame rate and is shown in Fig.2b in cylindrical coordinates for 5 selected frames (see Supplementary Note 3 and Supplementary Movie 1). Recordings over the initial 49,000 round-trips (~ 5 ms) in 8 consecutive reactors (out of 195) are plotted in Fig.2c and 2d, showing the soliton dynamics on finer time-scales.

The experiments indicate that the formation of a stable soliton-pair, resembling a molecule, generally requires multiple collisions. As shown in Fig.2b (Supplementary Movie 1) and 2c, while in some reactors[46] a stable soliton pair formed within 5 ms after only a few soliton collisions, in many other reactors hundreds of collisions were required. To examine the statistics of the synthesis, the cumulative collision numbers in all the reactors is plotted in Fig.2e against the total number of soliton molecules, measured at time intervals of 50 µs. The plot shows an approximately linear dependence, consistent with the collision theory of chemical kinetics which states that the rate of a gas-phase reaction is proportional to the collision frequency[46] (See Supplementary Note 4).

The group velocity difference between single solitons and soliton molecules[32–35] causes them to move to slightly different positions within their time-slots, while remaining trapped, as seen in Fig.2c; see also the reference slot in final state in Fig.2b. The optomechanical lattice is robust enough to host all the soliton reactions until they are completed, without destabilization.

By retrieving the soliton spacings and phases from the DFT signal[3], the complex nonlinear dynamics in hundreds of parallel soliton reactors can be recorded simultaneously over a long time period, revealing the stochastics of soliton-molecule formation (see Fig.3). Panel (i) – (iii) in Fig.3a (with corresponding trajectories in Figs.3b – 3e) show reaction processes in 3 parallel reactors in which the two solitons in each time-slot attempt to transit from a phase-uncorrelated long-range bound state (~60 ps spacing) to a phase-locked soliton-molecule (3.8 ps spacing and π-phase difference[47]) after the pump power is perturbed (same as in Fig.2, see more examples in Supplementary Note 4). While in panel (i) the formation process is completed within 1000 round-trips, following a rather simple trajectory (spacing and phase evolution in Fig.3b and interaction plane[16] in Fig.3c), the reaction shown in panel (ii) lasted more than 3000 cavity round-trips and followed a more complex trajectory (see Fig.3d). Panel (iii) shows the frequently observed case of strong interactions between two solitons lasting thousands of cavity round-trips without, however, giving birth to a soliton molecule (see Fig.3e). Instead, the two solitons strongly repel each other, before drifting towards the next collision. We observed that the soliton motion at most separations is stochastic, reminiscent of a one-dimensional random walk with fixed step-length[48]. This is probably caused by phase-dependent inter-soliton forces[39,49] that are constantly varying in strength and direction, weakly perturbed by non-solitonic components[45]. The initial phase difference between two solitons before binding is also random[8], accounting for the widely different trajectories from reactor to reactor. When the soliton separation is less than the molecular spacing (<3.8 ps), however, a strong repulsive force emerges[49] which quickly pushes the solitons apart. This is a ubiquitous feature not only in the synthesis of soliton-molecules but also in their dissociation, as described below (see Supplementary Note 4.)

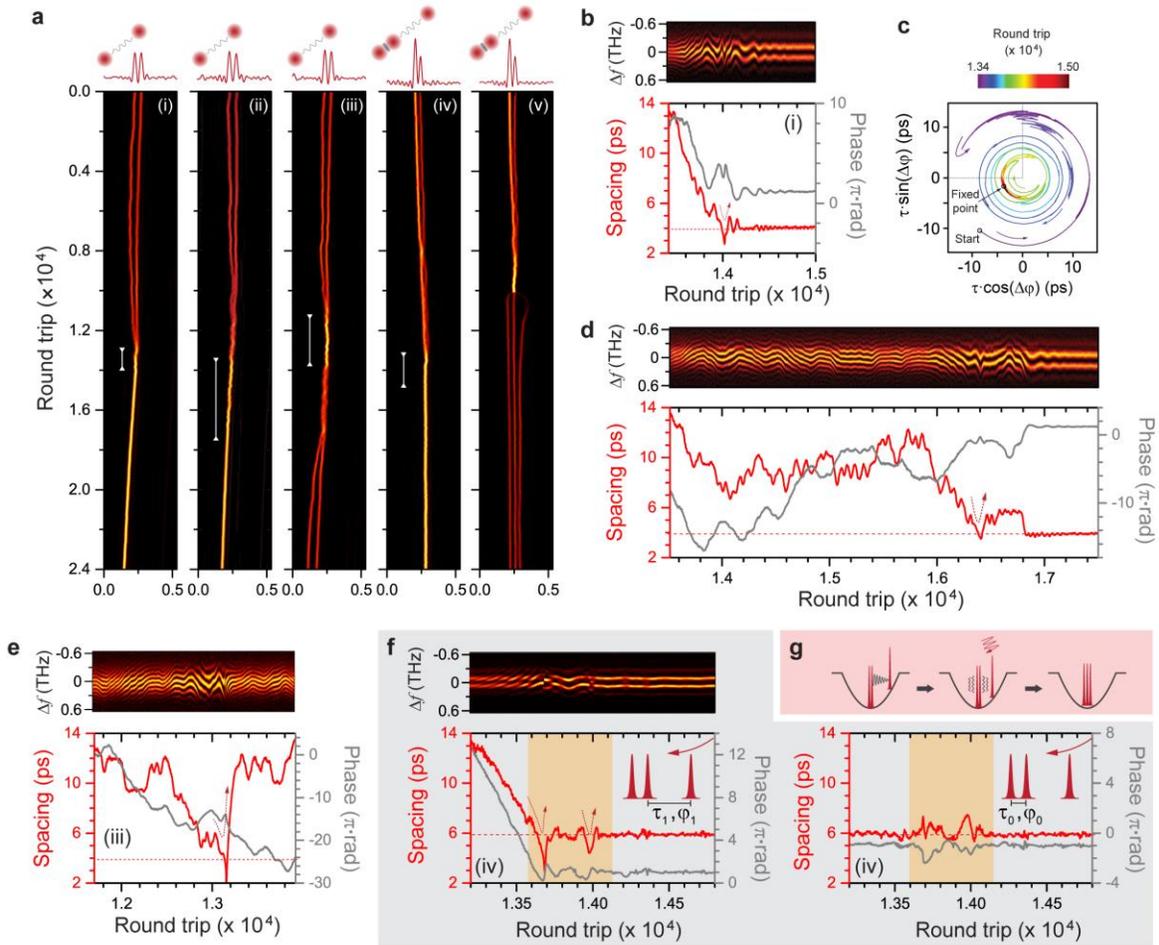

**Fig.3 | Dynamics of soliton-molecule synthesis. a** Panels (i)-(iii): time-domain recordings in three selected time-slots showing reactions between long-range bound solitons (~60 ps spacing) and soliton-molecules (~3.8 ps inner spacing). Panels (i) and (ii) show successful synthesis, while panel (iii) shows dissociation after collision. Panels (iv) and (v) show reactions between a single soliton and soliton-pair molecule (~50 ps inner spacing), resulting in formation of a soliton-triplet (~6 ps inner spacing, panel (iv)) and in repulsion between all three solitons (panel (v)). **b-e** DFT-signal (upper) and retrieved spacing and phase evolution (lower) for the ranges marked by white lines in panels (i)-(iii) (**c** is the interaction-plane plot of **b**). The horizontal dashed lines mark the spacing of a stable molecule. **f** DFT signal (top panel) and the retrieved spacing and phase relation between neighbouring solitons (lower panels) during soliton-triplet formation (panel (iv) in **a**). The horizontal dashed-lines mark the minimum soliton spacing (~6 ps) in the final state. **g** Schematic of soliton-triplet formation. A single soliton collides with a soliton-pair multiple times, disturbing the original molecular bond and resulting in formation of a soliton-triplet.

Soliton-triplets could also be formed by global control. We first prepared a soliton supramolecule in which most time-slots hosted a single soliton bound with a phase-locked soliton-pair, their velocities being synchronized by long-range interactions[8,50]. Then we abruptly increased the intra-cavity loss using the fast tunable attenuator, which weakened the dispersive-wave perturbation (See Supplementary Note 3). Consequently, attraction overcame repulsion, initiating soliton reactions. Two examples of three-soliton reactions are shown in panels (iv) and (v) of Fig.3a. In panel (iv), collision between the soliton pair and the single soliton resulted in formation of a phase-locked soliton triplet. The measured trajectories between neighbouring solitons during synthesis in panel (iv) are shown in Fig.3f. The single soliton collides strongly with the soliton-pair, resulting in strong disturbance to the soliton-pair (highlighted in yellow in Fig.3f) before the establishment of a second molecular bond (the reaction process is sketched in Fig.3g). In panel (v), however, similar collisions between a soliton pair and a single soliton result in dissociation of the soliton-pair molecule, followed by strong repulsion between all three solitons. These results show that three-soliton reactions in parallel reactions can be highly diverse. For more examples, see Supplementary Note 4.

**Dissociation of soliton-molecules under global control**

Phase-locked soliton molecules can also dissociate into single solitons under global control (Fig.4a). A typical example is recorded and plotted in cylindrical coordinates in Fig.4b for 5 selected frames (for full recording see Supplementary Movie 2). The reaction is initiated by a slight decrease in pump power, which enhances the dispersive-wave perturbations, causing rapid break-up of the soliton-molecules. This process is much faster than soliton-molecule formation, which generally requires multiple collisions. The dissociation follows highly diverse trajectories from reactor to reactor, as seen in Fig.4c (time-domain) and Fig.4d (DFT). We attribute the stochastic fluctuations during the early stages of dissociation to noise-like repulsive forces between the solitons, initiated by randomly excited dispersive waves[30,43]. After dissociation, long-range binding between the solitons is gradually established, eventually settling down after a few milliseconds (See Supplementary Note 5 for details).

To estimate the reaction rate during dissociation, we performed a statistical analysis based on defining criteria for determining the completion of a reaction: full dissociation for a separation of 14 ps or greater (the maximum spacing retrieved from DFT signal), and long-range soliton binding for a separation of 55 ps or greater. We plot the total number of soliton molecules and long-range double-solitons during dissociation against the number of round-trips (Fig.4e). Both measurements are roughly exponential, indicating that the dissociation rate is proportional to the number of un-dissociated reactants[46]. This allows us to estimate a soliton-molecule "half-life" of ~1200 round-trips (~120 µs) using the 14 ps criterion. (See Supplementary Note 5 for details.)

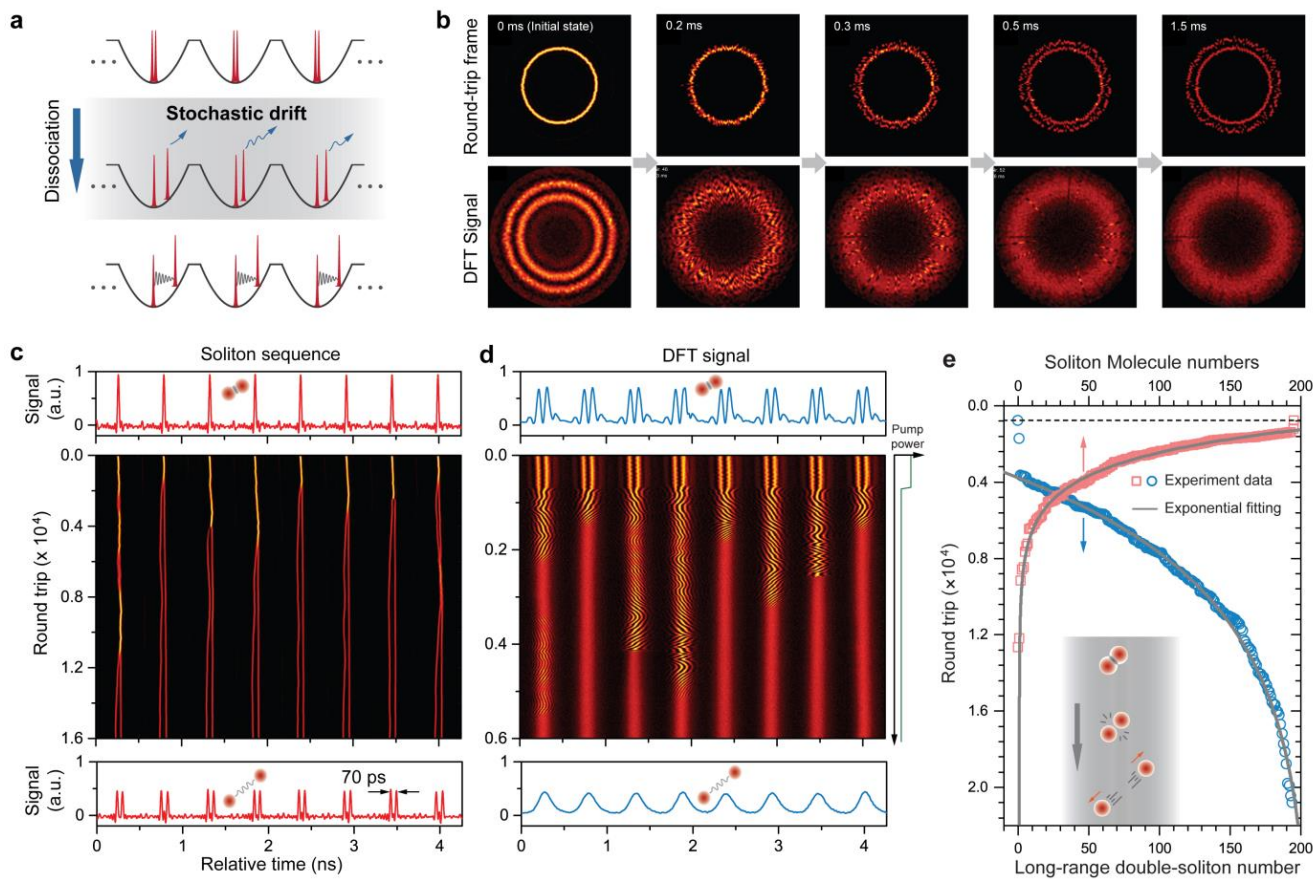

**Fig.4 | Dissociation of soliton molecules in parallel reactors. a** Schematic of the dissociation process in the case when all the reactors contain identical soliton molecules. By perturbing the pump power or the cavity loss, the molecular bonds collapse and the solitons start to drift stochastically and diverge, before reaching a stable long-range bound state. **b** Upper panels: selected frames from experimental measurements of the dissociation process in all 195 reactors, plotted in cylindrical coordinates (see Methods and Supplementary Note 2). Lower panels: the corresponding DFT signals. Initially all the time-slots are occupied by identical soliton molecules with spacings of ~3.8 ps and a relative phase of ~π. The gradual smearing-out of interferometric fringes in the DFT signal indicates dissociation. See Supplementary Movie 2 for the complete recording. **c**, **d** Time-domain sequence in 8 consecutive time-slots over the first 16000 round-trips (~1.8 ms) and the corresponding DFT signal, exhibiting instant collapse of the molecular bonds following pump-power perturbation. The dissociation trajectories are highly diverse, as indicated by fringes in **d**. Upper and lower panels in **c** and **d** show the signal traces for the initial and final round-trips. **e** Populations of soliton molecules (red squares) and long-range double-solitons (blue circles) over all 195 reactors, plotted against round-trip number during dissociation, fitted to exponential functions (grey curves). The horizontal dashed line marks the perturbation time.

The soliton motion during dissociation can be retrieved from the DFT signal (Fig.5). Panels (i)-(iii) in Fig.5a show three trajectories, initiated by perturbing the pump power. Panel (i) shows fast dissociation (<1000 round-trips) with a relatively smooth trajectory, as indicated by the retrieved spacing and phase (Fig.5b and c). In panel (ii) (DFT signal in Fig.5d), however, the trajectory is terminated after only a few round-trips by an abrupt drop in soliton spacing to below 2 ps, which immediately triggers strong repulsion between the solitons[49]. This is observed in many other reactors when the spacing falls below its initial value (~3.8 ps in this case), although occasionally the interacting solitons are extinguished (see Supplementary Note 5). Panel (iii) shows a dissociation lasting >10,000 round-trips with a random-walk-like trajectory (see Fig.5e). Other features are also evident, including strong repulsion for soliton spacings below the initial value, and metastable spacings of ~11 ps. For more details and examples see Supplementary Note 5.

Dissociation of soliton-triplets follows even more complex dynamics, as seen when the system is loaded with soliton-triplets in each reactor and then perturbed by decreasing the cavity-loss. Two examples are shown in panels (iv) and (v) of Fig.5a. In panel (iv) dissociation breaks only one triplet bond, leading to a long-range soliton pair bound to a single soliton. In panel (v), both molecular-bonds between the three solitons are severed, resulting in three phase-uncorrelated single solitons. (See details and more examples in Supplementary Note 5.)

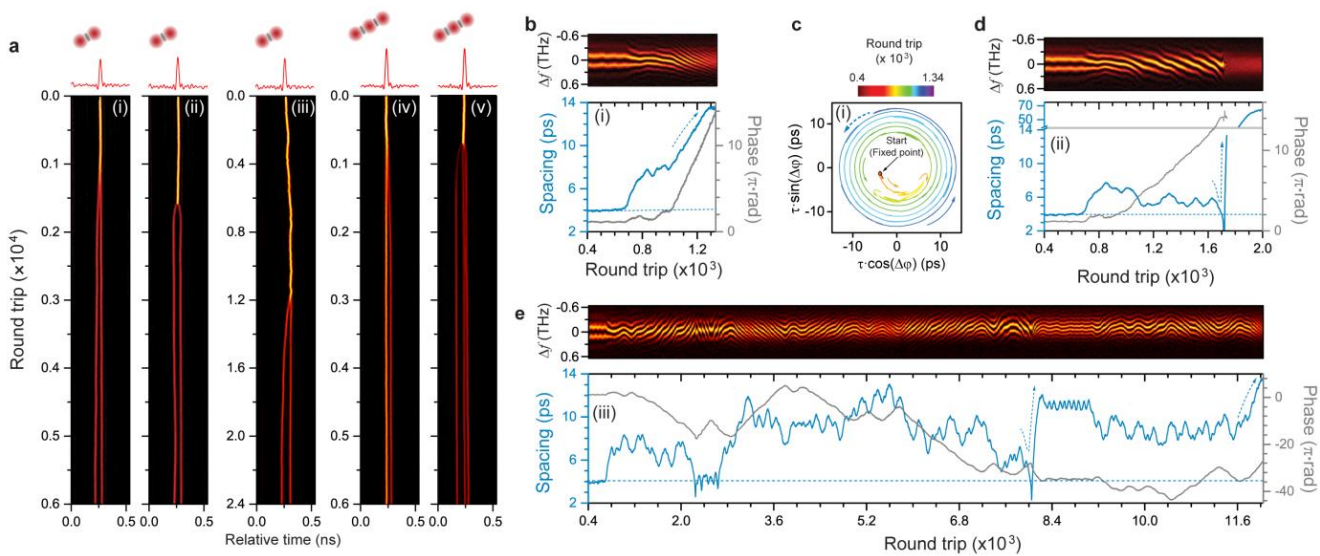

**Fig.5 | Dissociation dynamics of soliton molecules. a** Panels (i)-(iii): time-domain recordings of soliton-pair dissociation in three selected reactors (~3.8 ps inner spacing). Panel (i) shows a rapid dissociation with a rather smooth trajectory as seen in the DFT signal in **b** and plotted in the interaction plane in **c**. Panel (ii) shows a dissociation that ends with the abrupt repulsion of the two solitons after the spacing falls below ~ 3.8 ps, as retrieved from DFT signal shown in **d**. Panel (iii) shows a rather longer dissociation process in a different reactor, with a random-walk-like trajectory as shown in **e**. The horizontal dashed-lines in **b**, **d**, and **e** mark the soliton spacing in the initial state. Panel (iv) and (v) shows the dissociation of two soliton-triple molecules in parallel reactors, resulting in (iv) creation of a soliton-pair molecule and a single soliton, and (v) three uncorrelated single solitons (See Supplementary Note 5 for details and more examples).

## Selective all-optical control of individual reactors

Soliton interactions within selected reactors can be controlled by launching a sequence of precisely timed optical pulses into the laser cavity (Fig.1a)[10,24,42], permitting individual solitonic elements to be edited by XPM (see Methods and Supplementary Note 6). To demonstrate this, we first prepared a soliton supramolecule in which a mixture of long-range double-solitons and phase-locked soliton-pairs exist in the time-slots. To convert two long-range-bound solitons into a soliton molecule, we launched a train of ~200-ps pulses at the cavity round-trip frequency, precisely timed to interact with targeted time-slots over ~3000 round trips (see Fig.6a and 6b). The addressing pulse greatly enhances the attractive force between the solitons, leading to formation of stable soliton-molecules (see DFT signal and retrieved trajectory in Fig.6c). Note that typically many addressing pulses are required before soliton molecules form (See Supplementary Note 6 for details).

Conversely, to break apart soliton molecules in selected reactors, it is necessary to introduce a slight repetition rate offset, so that the XPM-induced forces vary continuously with time (see Fig.6d). By suitable choice of repetition-rate offset ($f_{ext} - f_{cav} \approx 20$ Hz, see Fig.6e), we achieved deterministic dissociation of selected soliton-molecules. The addressing pulses in this case operate like an optical "scissors", severing the molecular bond while traversing the target time-slot. The trajectories retrieved from the DFT signal are shown in Fig.6f, revealing that the traversing pulse does not simply pull the two solitons apart, but first compresses the soliton separation to below ~5 ps, triggering strong repulsion within 2~3 round trips, and eventually leading to the collapse of the molecular bond (similar behaviour is seen in Fig.4d. See also Supplementary Note 6).

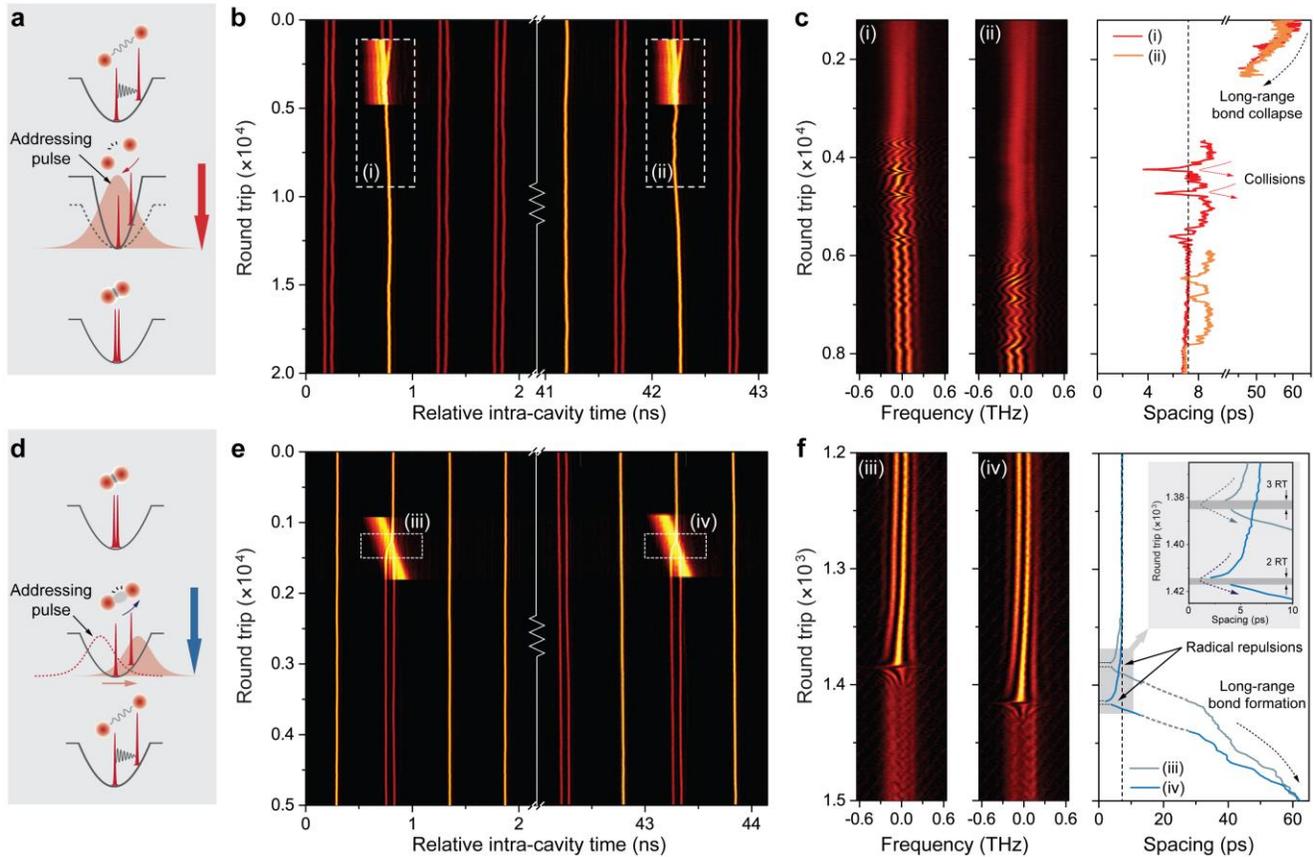

**Fig.6 | Control of reactions in selected time-slots by external pulses. a** An external addressing pulse overlaps continuously with an initially-prepared long-range soliton pair, causing strong attractive forces between them that result in formation of a soliton molecule. **b** Time-domain recordings of two processes in reactors hosting long-range soliton pairs. A 200-ps-long addressing pulse overlaps with the solitons over ~3000 round trips, causing the two solitons to drift together, interact, and form a soliton molecule. **c** DFT signals of the interaction within the white dashed boxes (i) and (ii) in **b**, together with the evolution of the inter-soliton spacing. **d** Schematic of soliton-molecule dissociation by an external addressing pulse. The addressing pulse traverses the soliton molecule, quickly severing the molecular bond. The two dissociated solitons become long-range bound in the end. **e** Time-domain recordings of the dissociation processes in reactors containing soliton-pairs. Addressing pulses with a slight repetition-rate offset "sweep" across the soliton molecules over 200 round trips, causing breakage of the soliton-molecule, followed by the establishment of long-range binding. **f** DFT signal for the processes within the white dashed boxes (iii) and (iv) in **e**, together with the evolution of the inter-soliton spacing. Strong repulsion is observed during dissociation when the spacing falls below ~5 ps, which occurred within 2~3 round trips, as shown in the inset.

**Discussion and conclusions**

Compared to platforms commonly used in the study of soliton dynamics, such as microresonators[12,13,41], passive fibre loops[29,42], and solid-state mode-locked lasers[3,10], optoacoustically mode-locked fibre lasers[8,24,36] offer several advantages. The fibre laser cavity is simple to configure, offering high flexibility in terms of gain, loss, nonlinearity and dispersion—critical parameters for controlling and studying optical solitons. The laser operation does not rely on external driving or feedback control, and the lasers self-start and run stably even when significantly perturbed. The fibre geometry allows easy incorporation of control elements and ports, the long cavity length permits accommodation of large numbers of solitons, and the optomechanical lattice ensures they are evenly spaced around the cavity—essential for the study of soliton interactions under controlled conditions.

Use of a 10 or 100 times longer laser cavity—realistic given low loss optical fibres— would permit a large increase in the number of parallel reactors (determined by the ratio between cavity round-trip time and acoustic cycle[24,37]) and support finer statistical studies. The factors determining the soliton reaction rate also require more investigation, in particular the stochastic motion during soliton interactions. The system could also be used to explore phenomena such as soliton explosions[51], pulsation[7,52], fragmentation[7,17], resonant vibrations[5,10], as well as the build-up[20] and extinction[18] of solitons.

Simultaneous control of multiple cavity parameters may allow more robust and diverse global control, further extending the capabilities of the system. All-optical control of solitons in individual time-slots may be further refined by adjusting parameters such as amplitude, repetition rate offset, and total overlap time differently in each reactor, so as to achieve "customized" editing of solitonic elements. As well as permitting a more extensive study of the multi-soliton dynamics, these improvements may lead to novel applications in all-optical information processing using solitons as data-bits[10,23,24,42].

**Acknowledgements:** The Max Planck Society is acknowledged for financial support.

**Author contributions:** The concept was proposed by W.H., M.P., and P.R., the experiments were carried out by W.H., D.H.Y., and J.H., the results were analysed by W.H. and M.P., and the paper was written by all authors.

**Competing interests:** Authors declare no competing interests.

**Additional information:** Supplementary Notes 1 – 6, Supplementary Movie 1 – 2.

## Methods

**Preparing soliton-supramolecule states.** In order to generate the desired soliton supramolecule state, we need to choose proper working point for the laser, adjusting the pump power level, cavity loss and most importantly carefully aligning the intra-cavity polarisation controllers. Sometimes dispersion compensating fibres are needed in the cavity to properly tailor the dispersive waves. In addition to the fast tunable optical attenuator, a manual tunable attenuator was also inserted in the cavity to introduce a loss-bias—crucial for finding a proper working point for the parallel reactions.

**Soliton sequences plotted in cylindrical coordinates.** The soliton sequence propagating in the laser ring-cavity follows a regular time-grid, each time-slot hosting one or more solitons. To simultaneously illustrate the soliton dynamics within all time-slots from frame to frame, we converted the temporal position $\tau_k(n)$ of the $k$-th soliton in the $n$-th time-slot into cylindrical coordinates following the relationship $(\rho_k(n), \phi_k(n)) = (\tau_0 + \tau_k(n), n2\pi/N)$, where $\tau_0$ is an arbitrary constant and $N = 195$ is the total number of time-slots (see Figs.2b and 4b). In each azimuthal "slice", the amplitudes of each soliton are indicated by a colormap.

**Diagnostic set-up.** We used a 33-GHz bandwidth photodiode for time-domain measurements, together with a 100 Gbits/s oscilloscope. This yielded a temporal resolution is ~15 ps, with a maximum recording time-span of 2.5 ms that was limited by the oscilloscope memory. To reach longer recording times (5 ms), in a few cases (e.g. in Figs.2b-2d and Fig.6b) we reduced the bandwidth to 16 GHz (50 Gbits/s sampling rate). For similar reasons the recordings in Fig.2b and Fig.4b were at different frame rates (5 kHz and 20 kHz), using the oscilloscope to take discrete shots of the round-trip signal during synthesis and dissociation. The DFT signal was obtained by linearly stretching the output pulse sequence through a 3-km-long SMF-28 with $\beta_2 = -22.5\text{ps}^2/\text{km}$, leading to a relative-frequency range of ~12 THz within each time-slot. The DFT signal is detected using a photodiode of 25-GHz bandwidth. The maximum soliton spacing that can be retrieved from the DFT signal is ~14 ps, limited by the bandwidth of the photodetector and the pulse stretching ratio.

**External addressing pulses.** The addressing pulses launched into the laser cavity were generated by modulating a single-wavelength laser at 1550 nm using a programmable pulse-sequence generator (~200 ps duration). The programmed pulse sequence followed a time grid that exactly matched the optomechanical lattice in the laser cavity, with 256 time-slots and repeated at ~7.344 MHz (laser-cavity length different from that under global control), in order to precisely overlap with selected time-slots. Ten evenly-spaced slots within the programmed time-grid were filled with addressing pulses, which were then amplified to ~20 W peak power in two amplifier stages. Launching of the addressing pulse was controlled by an optical switch (20-dB extinction ratio and 100 ns edge-time). The input port was a 50/50 coupler, which was also used as an output coupler. The polarisation state of the addressing pulses was adjusted so that they could be blocked by a polariser after exiting the cavity. In order to obtain a clean DFT signal without overlap from the addressing pulses, we inserted a 90/10 output coupler before the 50/50 output coupler.

**Data availability:** The data that support the plots within this paper and other findings of this study are available from the corresponding authors upon reasonable request.

**Code availability:** The code used in this paper is available from the corresponding author upon reasonable request.

# Supplementary Information for

# "Massively parallel optical-soliton reactors"

W. He *et al*.

# Supplementary Content

Supplementary Movie I: Soliton molecule synthesis in parallel reactors

Supplementary Movie II: Soliton molecule dissociation in parallel reactors

Supplementary References

# Supplementary Note 1: Experimental set-up

The setup used in our experiment, as sketched in Supplementary Figure 1, consists of four parts, including the optoacoustically mode-locked fibre laser cavity (Panel (i)), the set-up for controlling and varying the cavity parameters for global control (Panel (ii)), the set-up for generating external control pulses (Panel (iii)) for individual control, and the output-detection setup (Panel (iv)).

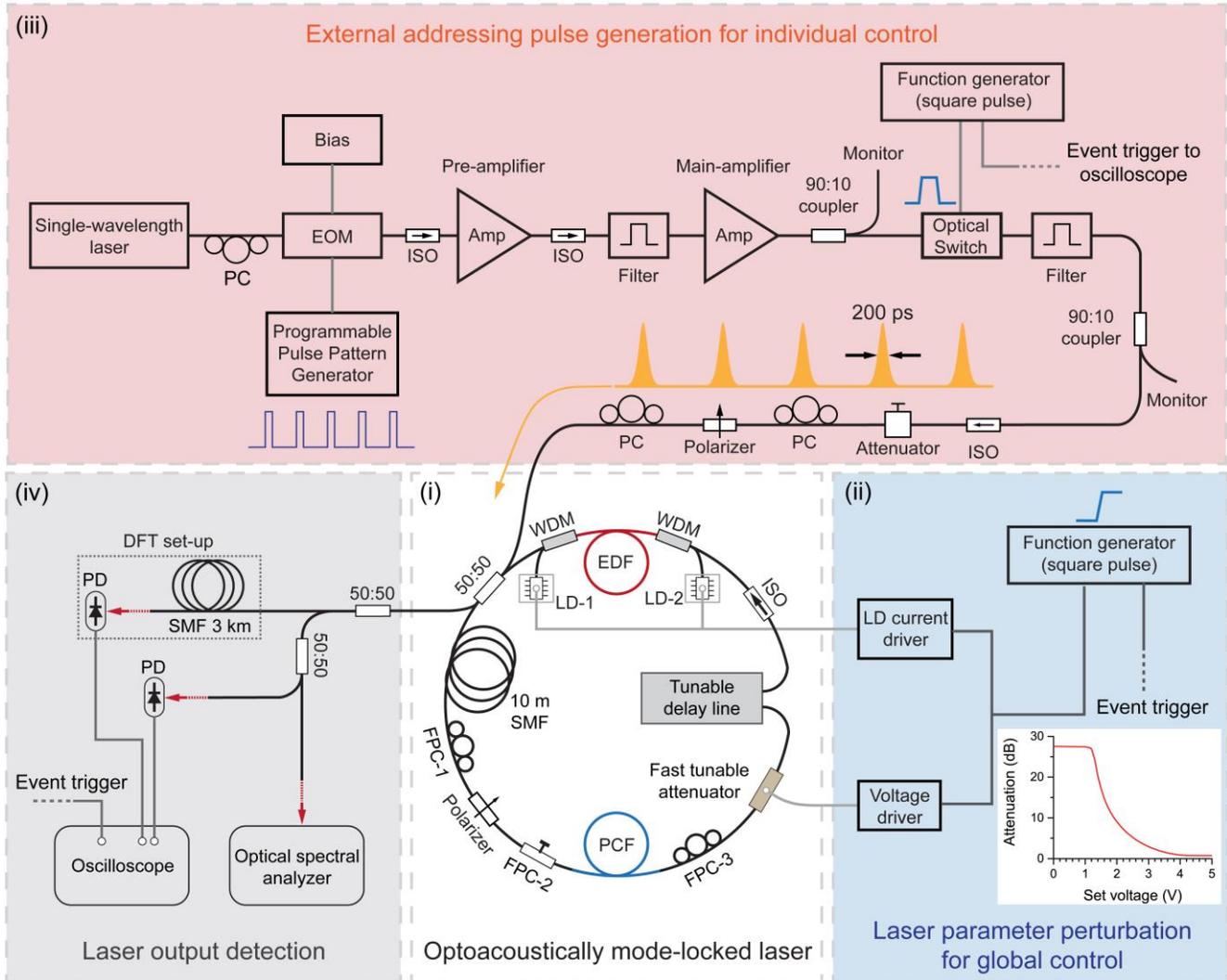

**Supplementary Figure 1 | Detailed sketch of experimental set-up.** The setup includes 4 connected parts, which are enclosed in 4 panels labelled as (i) – (iv) in the sketch. Panel (i): the optoacoustically mode-locked laser cavity which host the optomechanical lattice used as parallel reactors. Panel (ii): the global-control setup for perturbing laser parameters, including the EDF gain pump power (980 nm) through current drivers and the cavity loss through a fast tunable attenuator. Inset: The relation between the induced attenuation and the applied voltage upon the tunable attenuator. Panel (iii): the setup for generating external addressing pulses to enable individual-control. The addressing pulses are generated by modulating a single-wavelength laser and are launched into the mode-locked laser cavity through an coupler and interact with solitons in selected time-slot. Panel (iv): the diagnostic setup for recording laser output, including photodetector, oscilloscope and optical spectral analyzer. The DFT signal after temporal stretching using 3 km SMF-28 is recorded simultaneously. See text for details.

## The mode-locked ring-fibre laser

The optoacoustically mode-locked laser[1–3] has a ring-fibre cavity and consists of a gain section of erbium-doped fibre (EDF, 1.2-m length) pumped by two laser diodes at 980 nm at both forward and backward direction through the wavelength division multiplexers (WDMs), each with maximum 900 mW pump power. The inline isolator ensures the unidirectional operation of the laser cavity. The output coupler is a 50/50 coupler. A linear polariser and two fibre polarisation rotators (FPC-1 and -3) are used to initiate the mode-locking through nonlinear polarisation rotation (NPR)[4,5]. The optoacoustic mode-locking is enabled using a solid-core photonic crystal fibre (PCF)[6] with core diameter of ~1.9 µm, with the fundamental breathing acoustic mode ($R_{01}$-like) at 1.89 GHz and an optical birefringence of $1.5 \times 10^{-4}$ (group index). An additional FPC-2 is inserted before the PCF section for aligning the laser light with the principle axis of the PCF[1]. The insertion loss of the PCF section is 1.9 dB. A tunable delay line is used for adjusting the cavity length so as to allocate a specific harmonic of the cavity round-trip frequency into the acoustic resonance of the PCF. The total cavity length is around 20 m, which could be varied in different experiments, leading to different harmonic order. The results shown in Figs. 2 – 5 are based on a cavity with 195 time-slots (reactors), while for results in Fig.6, the cavity length increases to accommodate 256 time-slots (see Supplementary Note 6 for details).

**Global-control setup**

The global-control setup can apply fast perturbations upon laser cavity parameters, mainly the laser gain and cavity loss. The laser gain is controlled by changing the pump current of the pump-laser diodes in the EDFA section. A step-function change of the output voltage of a function generator is translated into pump current change (150 mA/V) while the laser is running, so as to perturb the interactions between the multi-solitons in all time-slots while the harmonic mode-locking state (i.e. the optomechanical lattice) remains stable. Another method for global control is to insert a fast variable optical attenuator (VOA402-3300, Boston Applied Technologies) in the laser cavity. The voltage step-change is first transformed into high-voltage by a transformer and then applied to the fast attenuator, with response time of 100 ns, and the relationship between the induced loss and the applied voltage (from the function generator) is shown as inset of Panel (ii). In addition, we found that the soliton reactions could also be initiated by changing the cavity length through the tunable delay-line, or by rotating the FPCs. However, due to the mechanical nature of the methods, the whole process lasts too long to be clearly recorded, while the final stages of the reactions are similar to those under pump and loss perturbation.

The global-control technique needs to be implemented in cooperation with a properly chosen working point of the laser cavity (mainly determined by the rotations of the FPCs, the pump power and the cavity loss). We insert a manual tunable attenuator in the cavity (not drawn) in order to set the loss "bias" for the laser cavity. In addition, the step-function voltage applied to the laser diode and the attenuator has a finite rise/fall time, which is also carefully adjusted in order to induce the steady reaction without destabilizing the optomechanical lattice. For example, in order to realize soliton molecule dissociation, the pump current is reduced by ~8%, and the fall edge of the voltage change was found to be optimized at 50 – 100 µs. If the pump current falls too fast, population oscillation would arise and eventually destabilizing the mode-locking state. Slower edge would cause, on the contrary, ultra-long dissociation time that may exceed the available time span of detection. Similar adjustments are also critical for cavity loss attenuations, which should be varied neither too fast nor too slow.

The descriptions on the individual-control setup (in Panel (iii)) is given in Supplementary Note 6 below, together with the working principles and some technical details.

**The diagnostic setup**

The time-domain pulse sequence are recorded using a 33-GHz photodetector (PD) and a 33-GHz oscilloscope (OSC). For continuous recording, the maximum recording length under 100 G/s sampling rate is 2.5 ms, while for some long lasting synthesis, the oscilloscope bandwidth would be downgraded to 16 GHz with 50 G/s sampling rate so as to reach 5-ms recording length (e.g. in Fig.2b – 2d). For frame-wised detections, an additional function generator would provide trigger signal to the oscilloscope, with 5 kHz-rate for synthesis and 20-kHz rate for dissociation so as to ensure full recording of the reaction (synthesis mostly lasts above 100 ms while dissociation mostly complete within 20 ms). An optical spectral analyzer (OSA) is also used to detect the averaged soliton spectrum under stationary states. The spectral resolution of the OSA is 0.01 nm.

The DFT signal is obtained by using a 3-km SMF-28 with a GVD of $-22.5$ $ps^2$/km ($D = +17.65$ ps/km/nm) and detected with a 25-GHz PD, corresponding to spectral resolution $\delta\lambda_{res} \approx 0.7$ nm at 1.55 µm wavelength, given by $\delta\lambda_{res} = 1/(B|D|z)$, in which $B$ is the analogue detection bandwidth, $D$ is the fibre dispersion, and the $z$ is the fibre length[7]. The maximum stretched duration is limited by the ~0.5-ns span of each time-slot without intervening neighboring slots, corresponding to a maximum bandwidth of $\Delta\nu$ ~12.5 THz or a $\Delta\lambda$ ~10 nm given by $\Delta\lambda = T/(|D|z)$, in which $T$ is the span of the time-slot[7]. For detecting the intimate spacing within the multi-soliton structures in each time-slot, the period of the interferometric fringes in the DFT signal needs to be determined out of numerical fitting. The maximum soliton spacing that can be retrieved from the DFT signal is ~14 ps, which is limited by the PD bandwidth and the stretching ratio. The precision of the retrieve soliton spacing is affected by the signal to noise ratio of the PD and the intrinsic timing jitter of the oscilloscope. Given ~2 ps uncertainty in the fringe period of interferometric DFT signal, the precision of the retrieved soliton spacing in our case should be ~0.1 ps (for soliton spacing around ~10 ps).

## Supplementary Note 2: Soliton sequences plotted in cylindrical coordinates

The recorded soliton sequence as output of the laser are segmented according to the cavity round-trip time and then plotted along each other so as to illustrate the dynamic changes of each soliton over consecutive round-trip in terms of their relative positions and amplitude within the cavity due to soliton interactions. We take for example a soliton supramolecule in which all time-slot consist of the same long-range double-pulse (ADS supramolecule[8]), while the entire optomechanical lattice consists of 160 time-slots. The time-domain pulse sequence of one round-trip recorded by the PD and OSC is shown in Supplementary Figure 2a, while the same pattern is precisely preserved after consecutive round-trip, as shown in Supplementary Figure 2b which have consecutive round-trip segment plotted along each other. Such "round-trip plot" is used frequently in the main article and also in the following supplementary figures.

The consecutive time-slots within one round-trip of the optoacoustically mode-locked fibre ring laser follow a regular time-grid determined by a fixed frequency of acoustic vibration in PCF-core. Therefore the pulse sequence within one round-trip can be segmented according to the span of a time-slot and then plot the consecutive time-slot segments in parallel in order to show the relative amplitude and positions of solitons over all time-slots, as shown in Supplementary Figure 2c with the example of the ADS supramolecule. We can also plot the consecutive time-slots in a special cylindrical coordinate, with radial distance of $\rho$ and azimuthal angle of $\phi$, which better illustrate the ring-cavity configuration that hosts the consecutive time-slots. The $2\pi$ azimuthal space is evenly divided into $N_{slot}$ divisions for the $N_{slot}$ consecutive time-slots in the fibre ring cavity. In each azimuthal "slice", the interacting pulses are plotted along radial axis, with their amplitude indicated by a colormap. Using an analytic description, the temporal position $\tau_k(n)$ of the $k$-th soliton in the $n$-th time-slot is plotted in the coordinate with its position following the relationship $(\rho_k(n), \phi_k(n)) = (\tau_0 + \tau_k(n), n2\pi/N)$, where $\tau_0$ is an arbitrary constant and $N$ is the total number of time-slots.

We plot the same ADS supramolecule in Supplementary Figure 2d, with 160 divisions, each occupying ~0.04 rad azimuthal space (highly exaggerated in the white-line marking), while the recorded signal within each time-slot are plotted along the radial axis within a span of one mechanical cycle (~0.532 ns in time or ~10 cm in space) marked by the white circle (not drawn in the main article). When a soliton sequence consists of soliton molecules, the limited bandwidth of the PD would translate the soliton-pair (or triplet) structure into a single pulse with doubled (or tripled) amplitude compared with that of a single-soliton. One example is shown in Supplementary Figure 2e, with the higher-amplitude indicating a soliton molecule. Such "time-slot plots" in cylindrical coordinate are used in Figs. 2b and 4b (and also Supplementary Movies 1 and 2) to illustrate the frame-wise recording of the synthesis and dissociation of the soliton molecules.

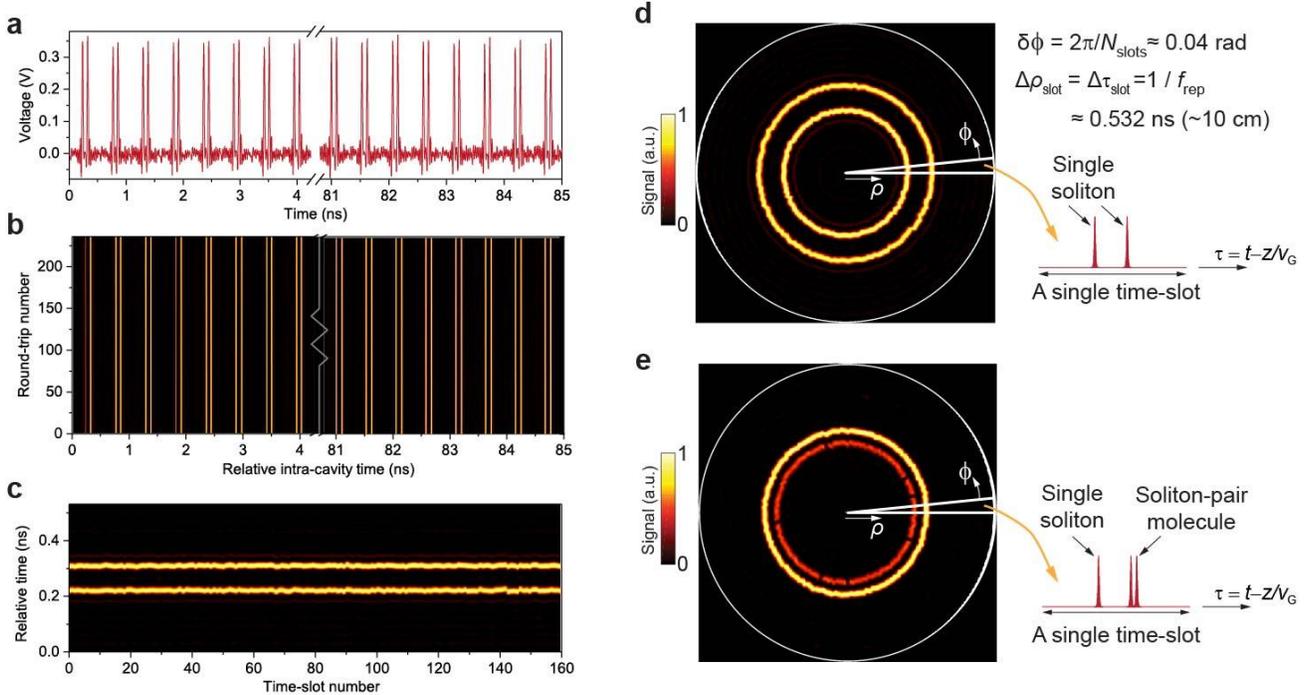

**Supplementary Figure 2** | **Different plot methods for the soliton sequence within a round-trip of the optoacoustically mode-locked fibre ring-laser**. **a** The recorded time-domain traces of an ADS supramolecule over one round trip. **b** The round-trip plot of the ADS supramolecule. **c** The time-slot plot of the ADS supramolecule. **d** The time-slot plot of the ADS supramolecule in a cylindrical coordinate, in which each time-slot occupied 0.04 rad azimuthal space and ~0.532 ns (or ~10 cm) radial space, marked by the white circle. **e** The time-slot plot of an soliton supramolecule that consists of both soliton-pair molecule and single-soliton in a cylindrical coordinate. The soliton molecule is recorded as a single pulse with twice the amplitude compared with that of the single-soliton.

## Supplementary Note 3: Optical spectra of soliton sequences

The optical spectra of the optoacoustically mode-locked laser measured using the OSA cannot follow fast changes in dynamic process as shown in the article. Nor can it distinguish the possible difference between the time-slots within one cavity round-trip. Nevertheless, when the initial states we prepared are stationary and homogeneous (i.e. all time-slots contain the same multi-soliton pattern), the measured optical spectra will give a finer structure (0.01 nm resolution) than that of the DFT signal out of pulse stretching (~0.7 nm). We present below the measured spectra for some of the prepared soliton supramolecules in order to illustrate some detailed features that cannot be resolved using the DFT signal.

**Two-soliton reactions**

The transition between the long-range bound double solitons and the phase-locked soliton-pairs are analysed in the article to study the soliton molecule synthesis and dissociation (Supplementary Figure 3a). The optical spectrum for the mode-locked laser in which all time-slots accommodate long-range double solitons are shown in Supplementary Figure 3b. The 3-dB bandwidth is measured to be 2.9 nm, corresponding to ~850 fs transform-limited pulse width, agreeing with the autocorrelation trace (inset of Supplementary Figure 3b). The spectral profile is mostly smooth with weak fringes appearing only in the vicinity of the dominant Kelly sideband, indicating that the two solitons have uncorrelated phase-relations while the binding is based on repulsive forces exerted by the dispersive waves (DWs, See Ref.[8]). After synthesis, all time-slots are filled with soliton-pair molecules, and the optical spectrum exhibit high-contrast interferometric fringes over the entire profile (as shown in Supplementary Figure 3d and 3e). The inner spacing between the constituting solitons can then be determined to be ~3.8 ps and their relative phase is locked at ~π.

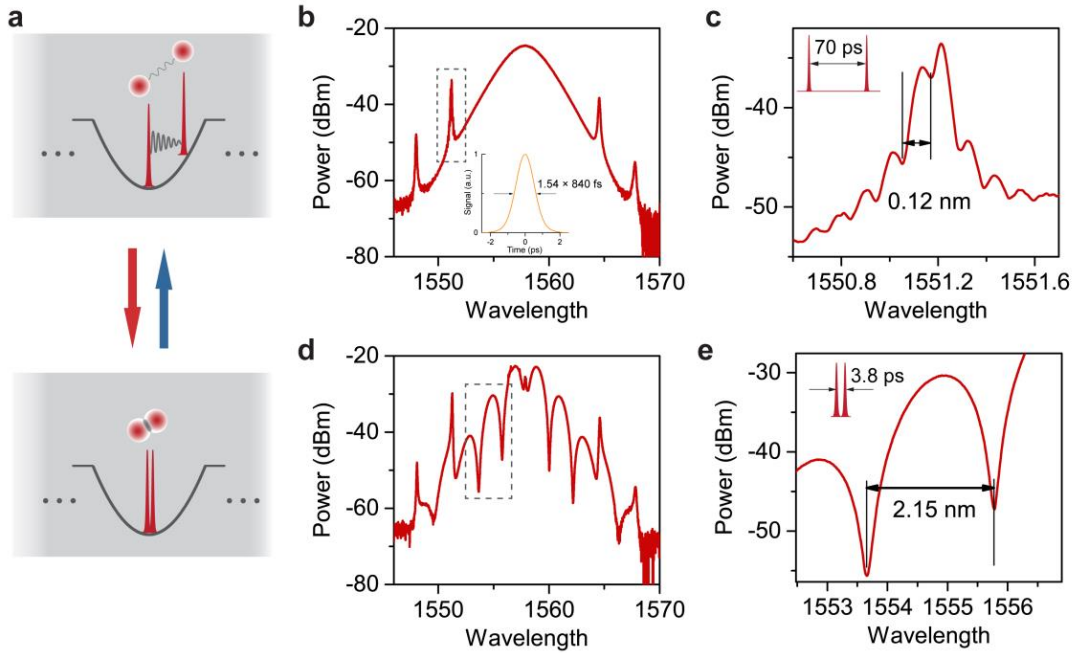

Supplementary Figure 3 | **Long-range double soliton and soliton-pair molecule. a** Transition between these two different multi-soliton state through synthesis and dissociation. Inset: autocorrelation traces of a typical soliton sequence, with retrieved duration of ~850 fs. **b**. Optical spectrum of the mode-locked laser with all time-slots filled with a long-range double-soliton, featuring smooth profile. **c**. Expanded view over the dominant Kelly sideband, featuring week fringes related to long-range spacing of 75 ps. **d**. Optical spectrum of the mode-locked laser, after the synthesis, with all time-slots filled with a soliton-pair molecule, featuring high-contrast fringe over the entire profile, indicating the fixed phase-relation (~π). **e**. The expanded view of the spectral fringe, with a period of 2.15 nm corresponding to 3.8 ps inner spacing.

The soliton-molecule spectrum is actually accompanied with a central CW peak (Supplementary Figure 3d), which are commonly seen in our experiments while it can hardly be recognized with DFT signal. This is due to the un-optimized working point of the FPCs after synthesis through pump power/cavity loss perturbation, which allows the co-existence of a CW background. Generally some slight adjustments of the FPCs afterwards can easily eliminate it.

**Three-soliton reactions**

A soliton-triplet molecule can be synthesized from a soliton-pair molecule and a single-soliton, and reversely, dissociated into them, as illustrated in the article (and depicted in Supplementary Figure 4a). The optical spectrum for the long-range bound state of soliton-pair molecules and single-solitons (soliton-pair/soliton) also exhibit fringes due to the phase-locked soliton-pairs, although with a lower contrast compared with that in Supplementary Figure 3d due to the coexistence of single-solitons. After the synthesis, the time-slots are mostly occupied with soliton-triplet molecules, with the corresponding optical spectrum shown in Supplementary Figure 4c. If the soliton-triplet molecule can be described in time domain as

$$E(t) = E(t+\tau_1)e^{i\varphi_1} + E(t) + E(t-\tau_2)e^{i\varphi_2},$$

where we assume all constituting solitons have the same envelope. Then the spectral interferogram of the complex envelope obtained from the Fourier transform should be

$$S(\omega) = |\tilde{E}(\omega)|^2 \{3 + 2\cos(\omega\tau_1 + \varphi_1) + 2\cos(\omega\tau_2 + \varphi_2) + 2\cos(\omega(\tau_1 + \tau_2) + \varphi_1 + \varphi_2)\}.$$

Accordingly, we can retrieve the inner spacings between the consecutive solitons as $\tau_1 = \tau_2 = 5.9$ ps and their relative phase $\varphi_1 = \varphi_2 \approx \pi$ within the soliton-triplet molecule from the measured spectrum shown in Supplementary Figure 4c which exhibiting multiple interleaving fringes.

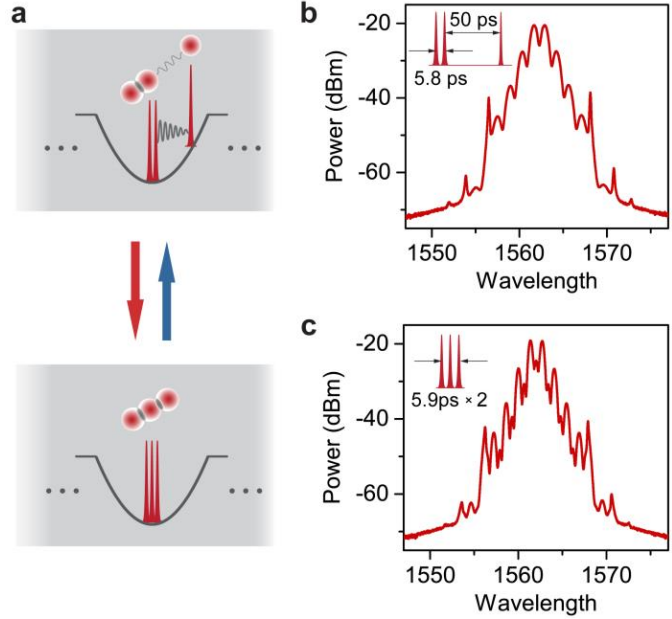

**Supplementary Figure 4 | Long-range bound soliton-pair/soliton and soliton-triplet molecule. a** Transition between long-range bound soliton-pair/soliton and a soliton-triplet molecule. **b** The optical spectrum of the mode-locked laser with all time-slots filled with long-range bound soliton-pair and single-soliton. **c** The optical spectrum of the same mode-locked laser, after synthesis, with all time-slots filled with a soliton-triplet molecule.

Note that the DFT signal of soliton-pair molecule and soliton-triplet molecule, when both with inner phase of π, have similar profiles in our experiments featuring two major peak symmetrically located about the central valley. Due to the limited span of each time-slot that can be used for dispersive pulse stretching, the weak bump at the centre of the soliton triple spectrum can hardly be recognized in the DFT signal. However, under similar bandwidth, the two major peaks in the DFT signal fringe of soliton triplet gives a significantly narrower width and higher amplitude than that of a soliton-pair, which can be noticed during the transition (See Fig.3f and Supplementary Figure 9 below).

Retrieving the spacings and relative phases between neighbouring solitons of a soliton triplet from the optical spectrum or DFT signal is however trickier than that of a soliton-pair, due to the increased number of fitting parameters and more importantly, the symmetric form of the interferogram that can give non-unique fitting parameters (flipping the value of $\tau_1$ and $\tau_2$, $\varphi_1$ and $\varphi_2$ would give the same interferogram). Especially, in case of radical collision and repulsions (as shown below in Supplementary Figure 10 and 13), we can no longer track the exact order the interacting solitons from the DFT signal and the consecutive spacing and phase-relation become uncertain. Therefore, only some simple, and smoothly changing dynamic of soliton triplets can be completely retrieved from the DFT signal (see Fig.3f in the main article and Supplementary Figure 18 below).

# Supplementary Note 4: Soliton molecule synthesis details

**Controlling long-range repulsion forces**

To prepare for the soliton-molecule synthesis, we make use of the recently discovered long-range bound state of solitons that rely on the balance between the repulsion force induced by the DW perturbation and the attraction force induced by the acoustic wave modulation[8]. In order to introduce soliton collisions that could lead to soliton-molecular bond formation, we can either reduce the repulsive forces or enhance the attraction force. In global-control method, we prefer to reduce the repulsion forces through varying the DWs (i.e. the sidebands) since they can be perturbed abruptly by changing the pump power and cavity loss, while the acoustic wave strength can hardly be varied equally fast that requires cavity-length tuning. According to our previous research, the repulsion forces depends on both the amplitude of the dominant sideband $A_d$, and the relative phase between the sideband and the perturbed soliton $\Delta\varphi_0$, which can be expressed in terms of carrier-frequency shift $\Delta\omega_d$ as below[8]:

$$\Delta\omega_d \propto A_d \exp(-h\Delta t)\psi(\Delta\varphi_0),$$

in which $h$ is a constant related to the bandwidth of the sideband (the decaying rate of the dispersive wave envelope), $\Delta t$ is the long-range spacing between the two solitons, and $\psi$ is a complicated oscillatory function that depends on the relative phase $\Delta\varphi_0$ as well as the dispersion and nonlinearity maps of the laser cavity.

In experiment we realized that a proper working point of the laser cavity, especially the states of the FPCs, are critical for achieving the desired soliton reactions in all parallel reactors. For the experiments of soliton-triplet formation as shown in Figs. 3a and 3f, we found a proper working point at which an abrupt increase of the cavity loss (~1 dB over 50-μs edge) can lead to attenuation of DW amplitude and result in collisions between the single-soliton and the soliton-pair molecule. The zoom-in figure of the dominant sidebands before and after this reaction is shown in Supplementary Figure 5a, from which the attenuated sideband amplitude can be readily noticed.

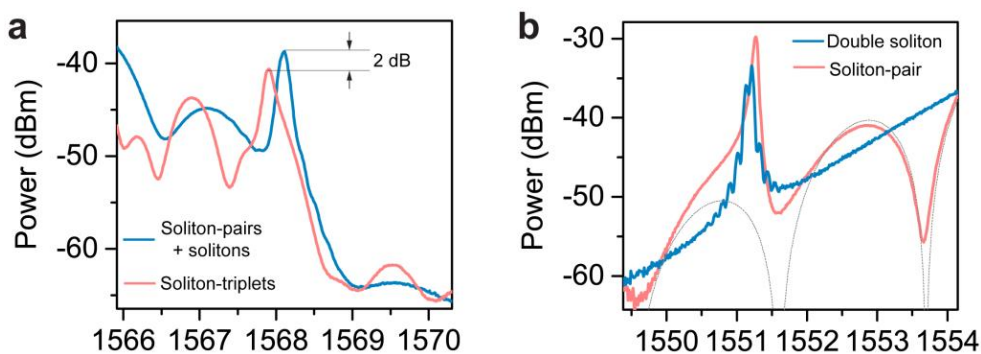

**Supplementary Figure 5** | **Comparisons of the dominant sidebands in soliton spectra during molecule synthesis** (spectra shown in Supplementary Figures 3 and 4). **a** Expanded view of the dominant sidebands for in laser spectra for the soliton-triplet molecule synthesis in parallel reactors. The initial states are long-range bound states of soliton-pair molecule and single-soliton (blue line). The final states are soliton-triplet molecules (red line). The transition is initiated with a 1-dB increase of the cavity loss which reduced the sideband power by ~2 dB, leading to subsequent soliton collisions. **b** Expanded view of the dominant sideband in laser spectra for soliton-pair molecule synthesis. The initial states are long-range bound double-solitons (blue line), and the final state is soliton-pair molecules (red-curve). The grey curve is the numerical fitting for the soliton-pair spectrum with sinusoidal modulation. The transition is initiated with an ~8% increase of the pump power, which shifts the working-point of the laser and destroys the fixed phase-relation between the DW and the perturbed soliton (indicated by the spectral fringe), leading to attenuated repulsion force, even though the peak magnitude of the sideband remains strong.

For the synthesis experiment of soliton-pair soliton molecule from long-range bound individual solitons, we found a working point at which a slight increase (~8%) of the pump power over ~100 µs edge could lead to almost full transition in all reactors. The varied pump power, instead of leading to soliton energy increase, actually leads to the disappearance of the locked phase-relation between the DW and the perturbed soliton, as well as the appearance of a noisy CW peak that might add up to the stochastic motion of the interacting soliton in all parallel reactors. Therefore, as we analysed the change of dominant sideband, we realized that it is the mainly the loss of phase-relation between the sideband and the perturbed soliton rather than a simple decrease of the sideband amplitude that leads to collapse of the repulsion force and consequently, the sliding motion of the two solitons towards each other (see Supplementary Movie 1). The exact impact of cavity parameter change upon this unique DW-soliton phase-relation are still not completely understood and remains an open question. Nevertheless, we were able to find a proper working point at which such long-range forces can be tailored to achieve the desired soliton reactions.

Soliton-pair molecule synthesis have also been achieved by abruptly increasing the cavity loss via the fast tunable attenuator. In experiment we prepare an ADS supramolecule with ~60 ps soliton spacing, and then increase the cavity loss by ~1 dB over a 50-µs edge. The experiment recording over part of the parallel reactors are shown in Supplementary Figure 6. Only a small portion of the parallel reactors see successful synthesis of soliton-pair molecules in the end probably due to the slight instabilities of the fast tunable attenuator, thus we did not choose this case for statistical studies in the main article.

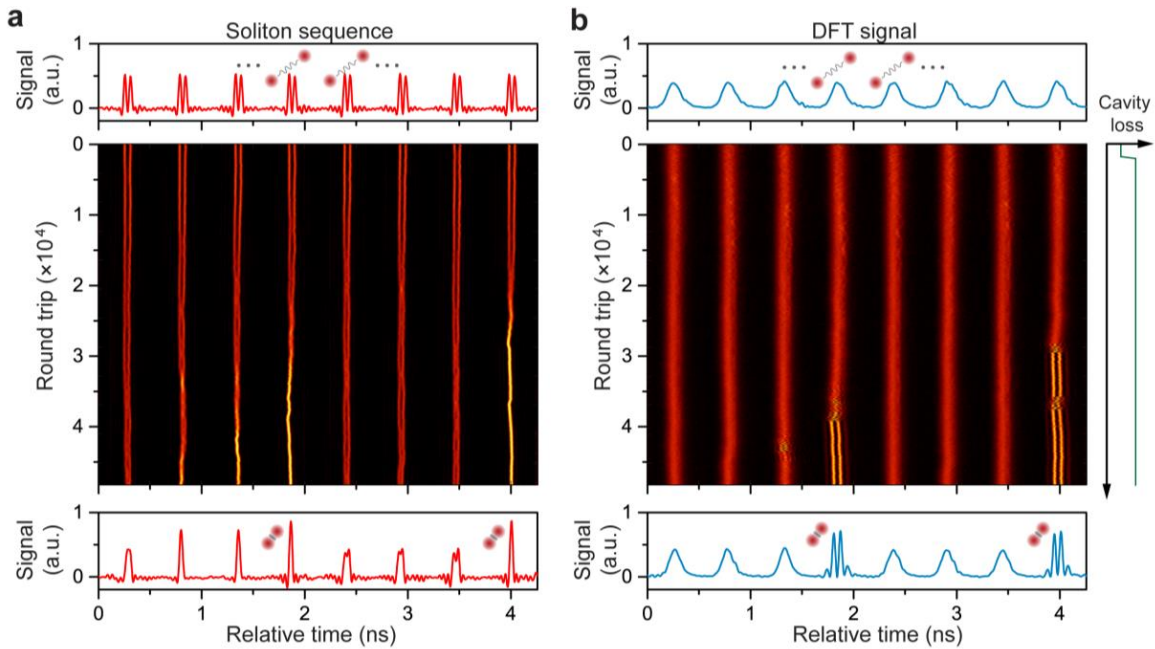

**Supplementary Figure 6 | Soliton-pair molecule synthesis via cavity-loss perturbation. a** Detailed recording over 8 consecutive time-slots over the initial 49000 round-trips (~5 ms). **b** The corresponding DFT signal. Only two time-slots see successful formation of soliton molecules as indicated by spectral fringe in **b**. Top and bottom traces in both panels are the signal in initial and final round-trips.

The instability in the spacing between the solitons at the initial long-range bound states, as can be noticed in the beginning of the Supplementary Movie 1, is actually a trade-off in setting the working point in order to achieve the desired soliton interaction. Although such instability can be easily mitigated by slight adjusting one of the FPCs (i.e. setting a proper polarisation bias for NPR action), however, a simple pump-power change under this status would not result in the aimed transition of all the multi-soliton state. Instead, it could leads to generation/elimination of some solitons that added up/moved from the supramolecular structure, as have been reported in our previous work (See Fig.5 in Ref[8]). The impact of the cavity parameters (or the working points) upon the long-range soliton forces still requires further investigations in order to explain the diversity in the responses of the soliton interactions. The optoacoustically mode-locked fibre laser provide an excellent platform for such investigation due to the constrained motion of solitons within each trapping potentials.

In order to improve the control technique for the soliton reactions, simultaneous control of multiple cavity parameters would be necessary, including e.g. pump power, cavity loss, and states of all FPCs, based on thorough understanding of the working points of the laser cavity. In one hand, the working-point setting for desired reactions would be less constrained and the reaction could probably be steadier. In the other hand, we could possibly achieve more complicated soliton reactions that can hardly reached by single-parameter control.

**Rate of soliton collision and soliton-molecule formation**

Resembling chemical reactions, the formation of soliton molecules requires multiple collisions of individual solitons in the trapping potentials before an effective collision occurred that leads to establishment of the molecular bond. Using the frame-by-frame recording of the long-term synthesis (as shown in Fig.2a and Supplementary Movie 1), we can roughly estimate the synthesis rate of the soliton molecule as well as collision rate of interacting soliton during the reaction. As shown in Fig.2c and 2d, due to the limited PD bandwidth, the separated peaks in each time-slot constantly merge into a single peak with higher amplitude and then repel each other again into two peaks. We regard one such action as one "collision" (as shown in Panel (iii) of Fig.3a). After many times of such collisions, they eventually merge into a stable soliton molecule (a single-peak signal) with stable fringe in the DFT single. We can then determine the completion of the reaction in each reactor. We repeat the synthesis experiment for several times, and plot the collision number as well as the soliton molecule number versus round-trip number (or time) during the synthesis. Two examples are shown in Supplementary Figure 7. We realized that although the collision rate and synthesis rate seems to vary from experiment to experiment, the cross plots reveal similar quasi-linear relationship between these two parameters, which uncovered a very interesting analogy to collision theory of the chemical kinetics that the rate of reaction is proportional to the collision frequency of the reactants. Although soliton reactions occurred between pre-sorted and paired solitons within the parallel reactors do not fully mirror the real chemical reactions in which all reactant freely collide with each other, the experiment result still provide interesting insight into the statistical properties of soliton interactions and can be further improved and modified to host other reaction conditions.

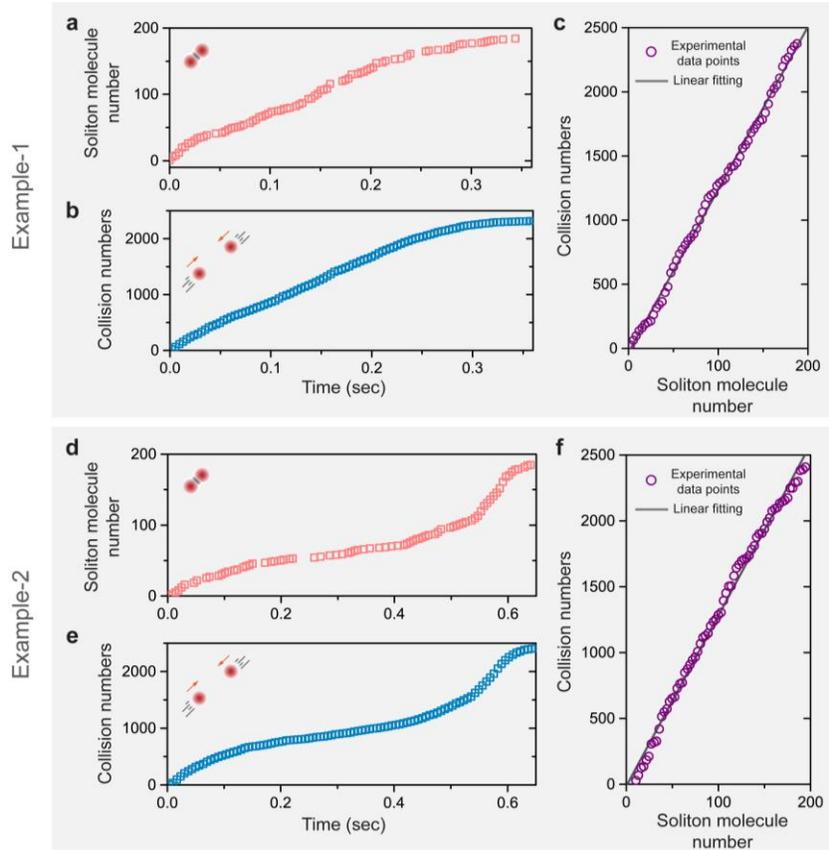

**Supplementary Figure 7** | **Correlation between synthesis rate and collision rate of soliton molecules**. Two examples of soliton-pair molecule synthesis experiments are shown under the same global-control method (~8% pump power change). **a**, **d** soliton molecule number over all parallel reactors that gradually increased over time during the synthesis. **b**, **e** Total soliton collision number over all parallel reactors. **c**, **f** cross-plots of collision number versus the soliton molecule number during each synthesis, both exhibiting quasi-linear proportionality.

The definition of one soliton-collision above did not consider the length of short-range interaction within each collision (i.e. the number of round trips during which the two solitons interact at short range. Compare Panel (iii) of Fig.3a in the main article and in Supplementary Figure 8a below). When considering the length of each collision and modified the definition accordingly, however, we found the cross-plot still exhibit quasi-linear proportionality. In the future, we expect to have finer estimation of soliton collisions with increased temporal resolution, which can further improve this statistical analysis.

The exact factors that influence the synthesis rate are still not completely clear. In addition to the working point, the perturbation strength of the cavity parameter are also found to influence the synthesis rate. In a few cases, we also observe segmented linear dependence (i.e. weak rolling points) in the cross-plot, which might result from weak inter time-slot influences due to un-optimized working point. As more and more soliton-molecules are formed, the time-slots with unfinished synthesis might be influenced by a varying background waves influenced by other finished time-slots. Another possible reason might be a random drift in some cavity parameters (FPC rotation states, environment temperature, pump-power instability, etc.) during the long-term synthesis. The exact mechanism for such collision-synthesis still requires detailed investigations, and the parallel-reactors scheme provide an ideal platform for further unfolding the mechanism with its statistical dimension.

**More examples of synthesis dynamics for soliton-pair molecules**

In the main article, a few examples of synthesis dynamics for soliton-pair molecules are illustrated (Fig.3a – 3e), exhibiting highly-diverse evolution trajectories. Here we give a few more examples recorded in the parallel reactors during the same synthesis (with pump-power variation) in Supplementary Figure 8. Panel (i) and (ii) in Figure 8a shows yet another two examples of successful formation of soliton-pair molecule within the initial ~2.5 ms, with the related DFT signal and retrieved spacing/phase trajectories given in Figures 10b and c. We can notice the gradually damped oscillation in the relative phase from the interferometric fringes in the DFT signal before soliton-molecular bond completely settle down. The same feature also appears in the examples in the article, indicating that the trapping potential formed by the short-range forces include some damping mechanism that can stabilize the molecular bond. Panel (iii) gives an unsuccessful collision with random-walk-like trajectory as shown in the DFT signal in Figure 8d. Just like most other cases, this soliton collision did not give birth to a soliton molecule despite the long interaction length (compared with the case in Fig.3a Panel (iii)).

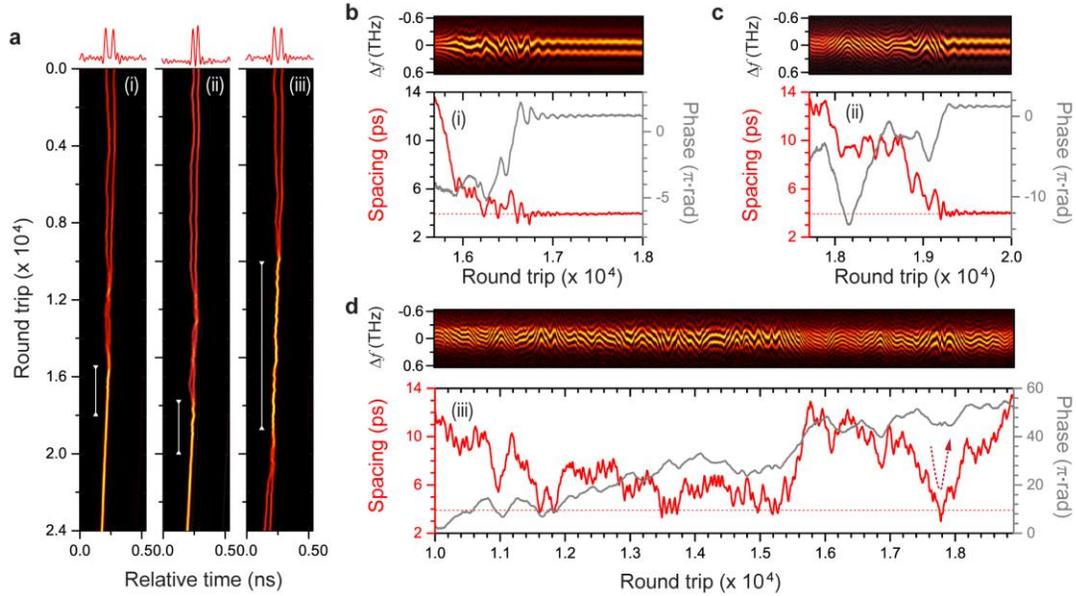

**Supplementary Figure 8** | **More examples of synthesis dynamics for soliton-pair molecules**. **a** Time-domain recordings of soliton interactions within 3 selected parallel reactors. In Panel (i) and (ii), the soliton collisions result in formation of soliton-pair molecule, with DFT signal and retrieved trajectories shown in **b** and **c**. In Panel (iii) the soliton collisions result in repelling of the interacting soliton in the end. The corresponding DFT signal is given in **d**, exhibiting random-walk like trajectories. The horizontal dashed-lines mark the soliton-molecular spacing at the final state.

## More example of synthesis dynamics for soliton-triplets molecules

Synthesis of soliton-triplet molecules are realized in the parallel-reactors scheme using global-control method (cavity-loss change) as shown in Fig.3 of the main article. Here we provide some details of this process. The synthesis starts with long-range bound state of a soliton-pair molecule and a single-soliton. After the cavity-loss is abruptly increased by ~1 dB over a 50-µs edge, the long-range repulsion collapsed and the single-soliton went through multiple collisions with the soliton-pair in each reactor, before an effective collision occurred and led to formation of a soliton-triplet. The synthesis in 8 consecutive time-slots (out of 160) recorded over the initial 2.5 ms are shown in Supplementary Figure 9b and 9c with both time-domain sequence and the DFT signal. We can first notice that the trajectories are highly diverse from reactor to reactor with both successful and failed synthesis cases. We can also notice the discrepancies in group velocities between different solitonic elements during the dynamic process (comparing with the reference time-slots that contain only a soliton-pair or single-soliton). The trapping potentials provided by the optomechanical lattice passively synchronize all these different elements and settle the velocity discrepancies into different trapping positions within each time-slot, avoiding "cross-talks" between the reactors.

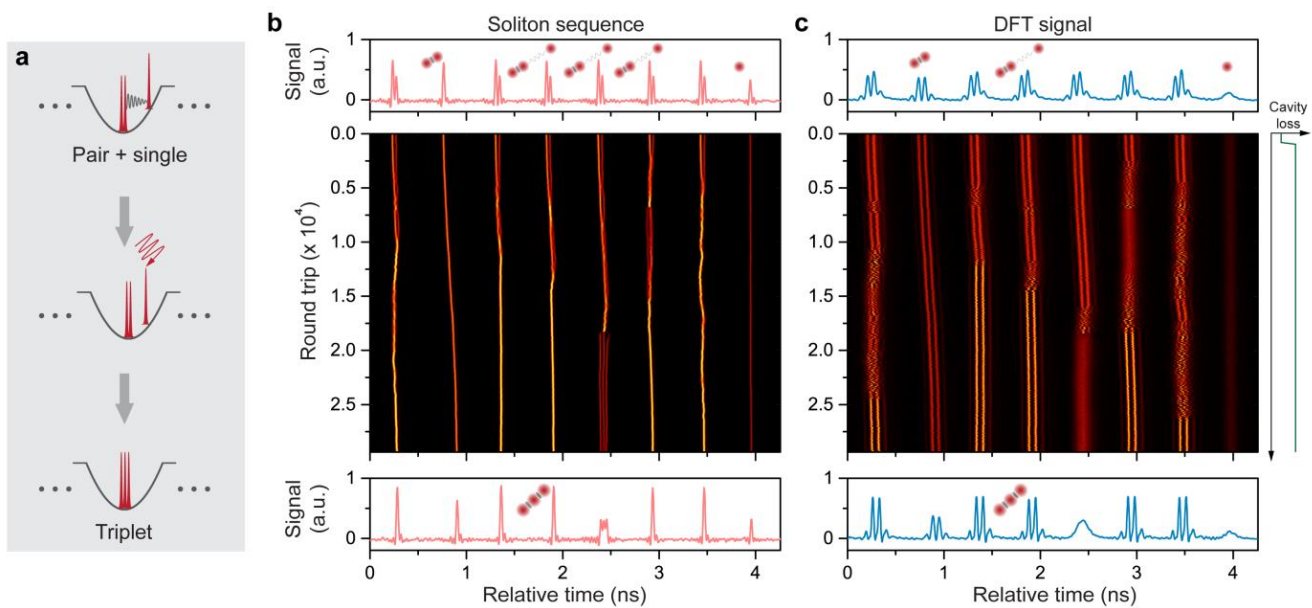

**Supplementary Figure 9 | Synthesis of soliton-triplet molecules in parallel reactors. a** Schematic of synthesis dynamics. Starting from long-range bound soliton-pair and single-soliton in each reactor, we reduce the long-range repulsion and cause collisions between the two solitonic elements, and eventually creating a soliton-triplet molecule. **b** The time-domain soliton sequence in 8 (out of 160) consecutive time-slots recorded over the initial ~29000 round-trips (2.5 ms). An abrupt 1-dB increase of cavity loss over a 50-µs edge is applied at the beginning to initiate the reactions. **c** The corresponding DFT signal of the time-domain sequence in **b**.

The interaction between the soliton-pair molecule and single-soliton in each reactor exhibit far richer diversity than what have been shown in Fig.3 of the article, and can be fully unfolded using the parallel-reactor scheme. We plot 8 examples out of the 160 reactors during the synthesis in Supplementary Figure 10. The top 8 panels are time-domain sequences and the bottom ones are the corresponding DFT signal over the interaction region. Panel (i) and (ii) shows successful synthesis of soliton triplets, with however highly different transition dynamics and interaction length. Panel (iii) shows an unsuccessful collision, after which the single-soliton is repelled away from the soliton-pair. Panel (iv) shows a rather interesting case, in which the two different solitonic elements swap their relative position after a long and rather complicated interaction. Detailed analysis over the DFT signal shows that the soliton-pair actually collapse during the interaction, then the three solitons recombine into this different pattern. We also notice that the long-range spacing between the two solitonic elements increased a lot compared with the initial case probably due to velocity discrepancies and varied long-range repulsion forces[8]. Panel (v) and (vi) show two examples in which the collisions results in repulsions and divergent motion between all three solitons, with different repulsion strength. The DFT signal below revealed the fact that the radical repulsion actually occurred within a very short time (a few round-trips, marked by while arrows in the DFT-signal plot) when the three solitons reached extremely intimate spacing. Panel (vii) and (viii) show two extreme cases in which the radical repulsions after soliton collisions result in extinctions of one or two solitons. From the DFT signal we can revealed that the extinct solitons have significant velocity difference with other single-soliton, probably due to shifted carrier-frequency. Consequently, the solitons no longer see balanced gain and loss due to limit gain bandwidth and are quickly eliminated after the repulsion. Similar repulsions are also observed during the dissociation, as shown in Fig.5d in the article and in Supplementary Note 5 below. Note that due to the limited PD bandwidths and stretching ratio, as well as non-unique fitting parameters, we were not able to retrieve faithful trajectory (phase and spacing evolution) for most three-soliton reactions except for some simple and smooth cases as in Fig.3g in the article. Improved detection range and fitting algorithm would be required for detailed investigations.

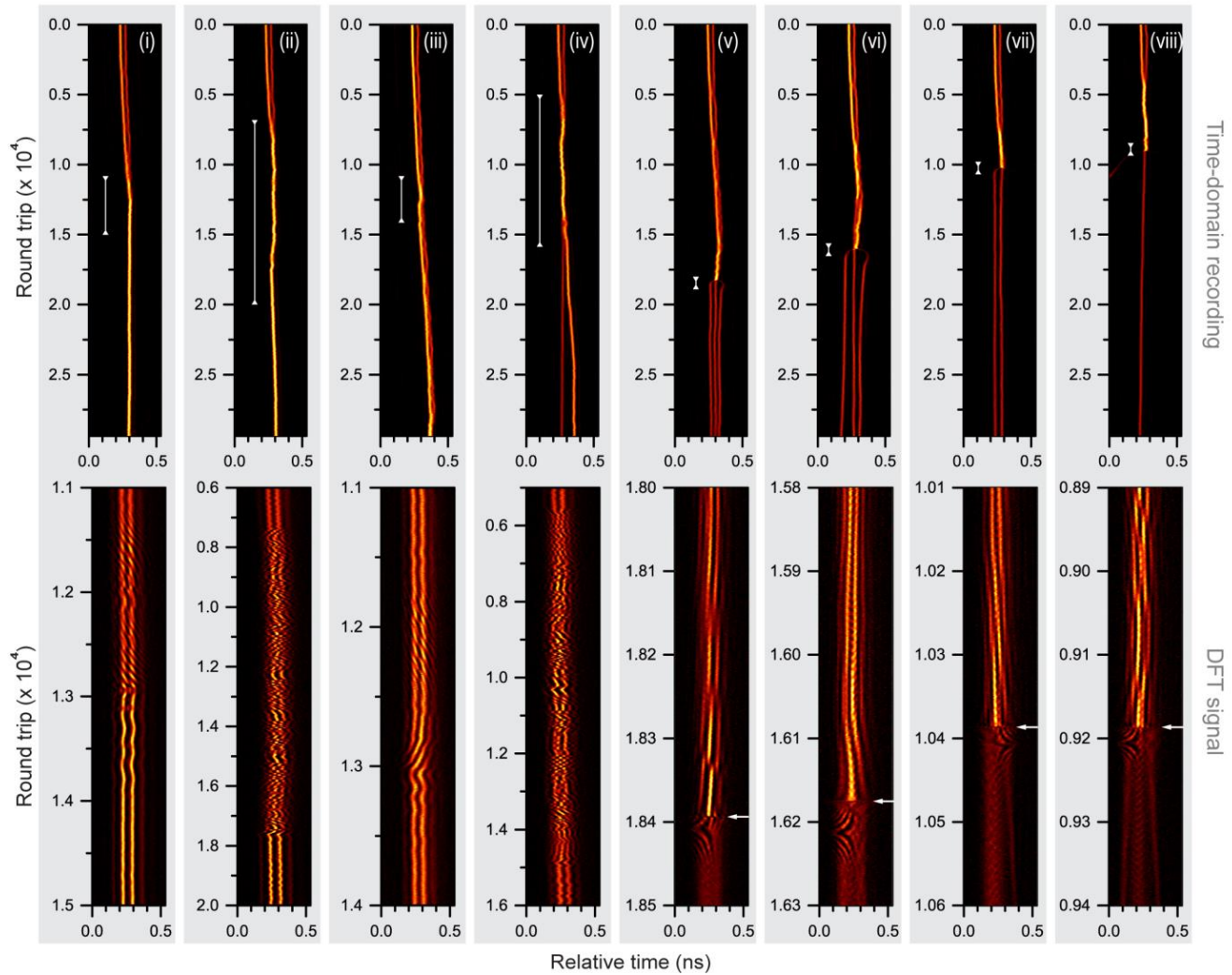

**Supplementary Figure 10 | Various collision dynamics for soliton-triplet molecules synthesis in parallel reactors**. The evolution trajectories in 8 (out of 160) time-slots are plotted. Top panels are time-domain recordings of individual time-slots over the initial 2.5 ms (~29000 round-trips). Bottom panels are the corresponding DFT signal over selected round-trip ranges marked by the while lines in top panels. All time-slots start with a long-range bound soliton-pair/soliton. Their collisions are initiated by 1-dB increase of the cavity-loss. See text for detailed descriptions.

## Supplementary Note 5: Soliton molecule dissociation details

### Control of long-range forces

The dissociation can be initiated simply by reversely perturbing one of the cavity parameter by which the repulsion forces can be enhanced. The examples of soliton-molecule dissociation given in Fig. 4 are the reversed pump-power change in the example of synthesis given in Fig. 2. In this case, the dispersive waves are re-established to exert strong repulsion forces that break up the soliton-molecular bonds. In a similar mechanism, the dissociation can also be initiated by decreasing the cavity-loss, through which the dispersive wave are directly enhanced and consequently lead to soliton dissociation. One such example is given in Supplementary Figure 11, which also exhibit highly diverse trajectories. Note that in most cases, the dissociated solitons in each time-slot are firstly repelled to different long-range spacings, which will eventually settle down to the same balanced spacings after a much longer time. A single-soliton time-slot is included in this plot as a reference for determining the relative position of the solitonic elements in the cavity.

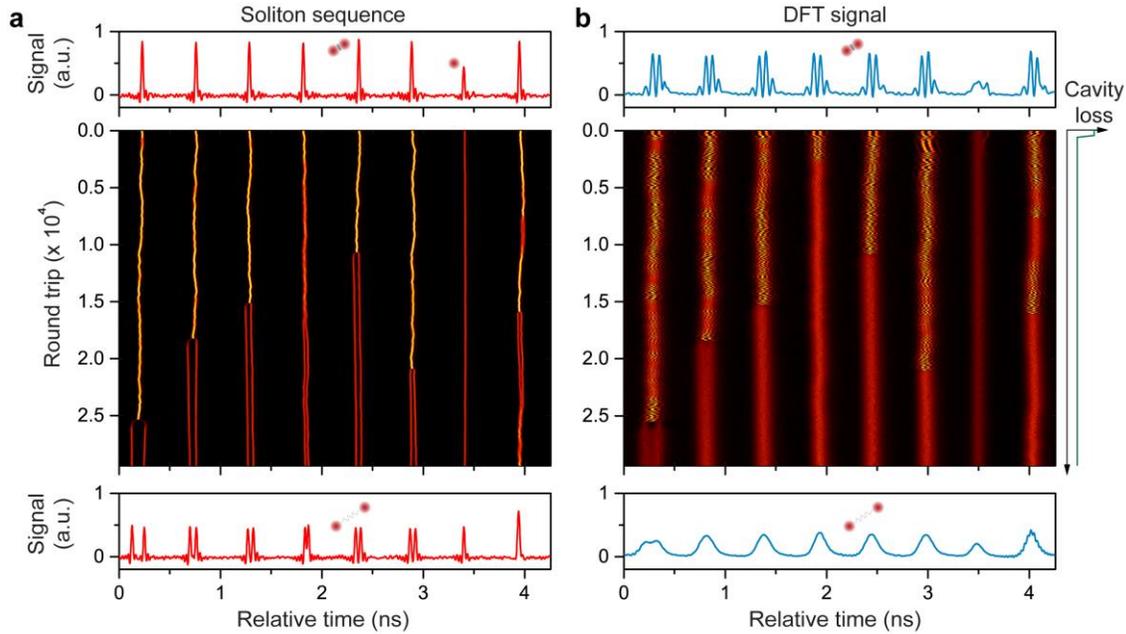

**Supplementary Figure 11 | Dissociation of soliton-pair molecule initiated by cavity loss control**. We prepare soliton-pair molecules with π-phase difference and ~5.9 ps spacing in the parallel reactors, and then decrease the cavity loss by ~1 dB over a 50-µs edge. **a** The soliton sequence in 8 (out of 160) consecutive time-slots over the initial ~29000 round trips of the dissociation. **b** The corresponding DFT signal.

### Dissociation rate of soliton-pair molecules

In the main article, a soliton molecule is regarded as dissociated when the two solitons are repelled beyond a significantly larger spacing than the molecular spacing. The exact spacing criterion is actually rather arbitrary within a reasonable range. We used a criterion of 14 ps in the article, which is the maximum resolution of the DFT diagnostic setup in our experiment. We also checked the statistical results under different spacing criteria and plot them in Supplementary Figure 12a. We can note that, since the initial repelling is generally fast, the exact decaying curve varied only slightly under different criterion. On the contrary, since the long-range spacing change becomes rather slow, the long-range double-soliton formation rate varied significantly under different criterion, as shown in Supplementary Figure 12b. Under all these different criterions, the results remains roughly to follow an exponentially-decay profile, which indicate that the dissociation follows the first-order reaction model described by the following rate equation:

$$\frac{dN}{dt} = -kN,$$

where $N$ is the instantaneous molecule number during the reaction at time $t$. The general solution is an exponential decay $N = N_0 \exp(-kt)$, with the "half-life" of $\ln(2)/k$.

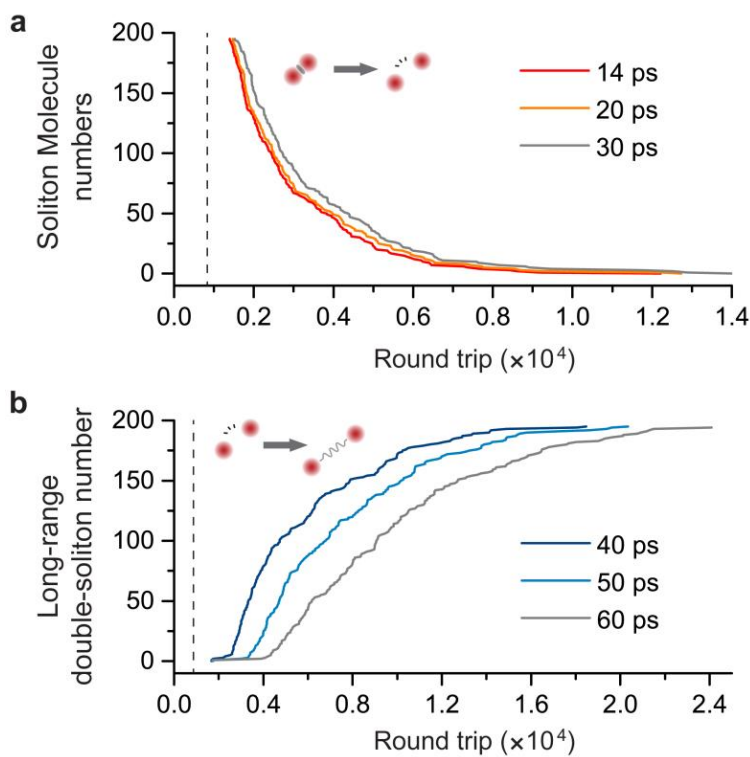

**Supplementary Figure 12** | **a** Soliton molecule number decaying and **b** long-range double-soliton number increasing over consecutive round trips under different spacing criterion. Different curve colors are used to mark counting results under different spacing-criterions. The vertical dashed lines in both plots indicate the time at which the perturbation occurred.

**Detailed dynamics of soliton-molecule dissociation**

Since the dissociation of soliton molecules occurs immediately after the perturbation, being significantly faster than the synthesis, we were able to observe massive events over all reactors with high-resolution signal (round-trip by round-trip, see Fig. 4c, 4d in the article and Supplementary Figure 11) within our recording capacities. Based on these massive observations, we have summarized a few primary features of the soliton interaction during as listed below:

(1) As the most obvious feature, the dissociation trajectories are highly stochastic and differ from time-slot to time-slot. This feature has been extensively illustrated in figures above and in the main article.

(2) Within the dissociation, radical repulsions between the two solitons always occurred when the soliton spacing narrowed down below the initial soliton-molecular spacing, as shown in Fig.5d and 5e in the article.

(3) There exist some metastable states at spacing other than original molecular spacing, in which the two solitons can temporally reside during their interactions.

(4) The stochastic motions of the soliton spacing other than the regions of radical repulsion and metastable states resemble one-dimension random walk with fixed step-length.

These features are also shared during the synthesis, although more examples can be obtained from the recording of the dissociation. We provide below additional examples for illustrating the features (2) – (4) and briefly discuss some details concerning each feature.

*Radical repulsions during dissociation.* In Supplementary Figure 13 we present six examples from the parallel reactors in which strong repulsions occurred during the soliton-pair molecule dissociation. (Examples in **a** and **b** use soliton molecules with 4.9-ps spacing while **c** – **f** use soliton molecule with 5.9-ps spacing. These summarized features are however found to be irrelevant to the exact parameter of soliton molecules). In all these examples, the time-domain recording, the DFT-signal, and the retrieved soliton spacing/phase are presented for each individual time-slot (highlighted within the yellow boxes).

The examples in **a** – **c** shows the dissociations that are terminated by soliton repulsions. During the random-walk-like motion, the soliton spacing could occasionally reached below the original molecular-bond spacing. The two solitons would then suddenly collide into an ultra-narrow spacing, followed a strong repulsion that pushes them away from each other. The "strength" of the repulsion can be revealed by the diverging velocity of the two solitons away from each other after the repulsion. The repulsion strength are rather weak in **a** while strong in **b** and **c**. The radical repulsion seems to strongly shift the carrier frequency of the two solitons oppositely, leading to the diverging motion of the two solitons. Due to the trapping effects of individual reactors, the diverging motions are then gradually mitigated and long-range double-soliton states are established.

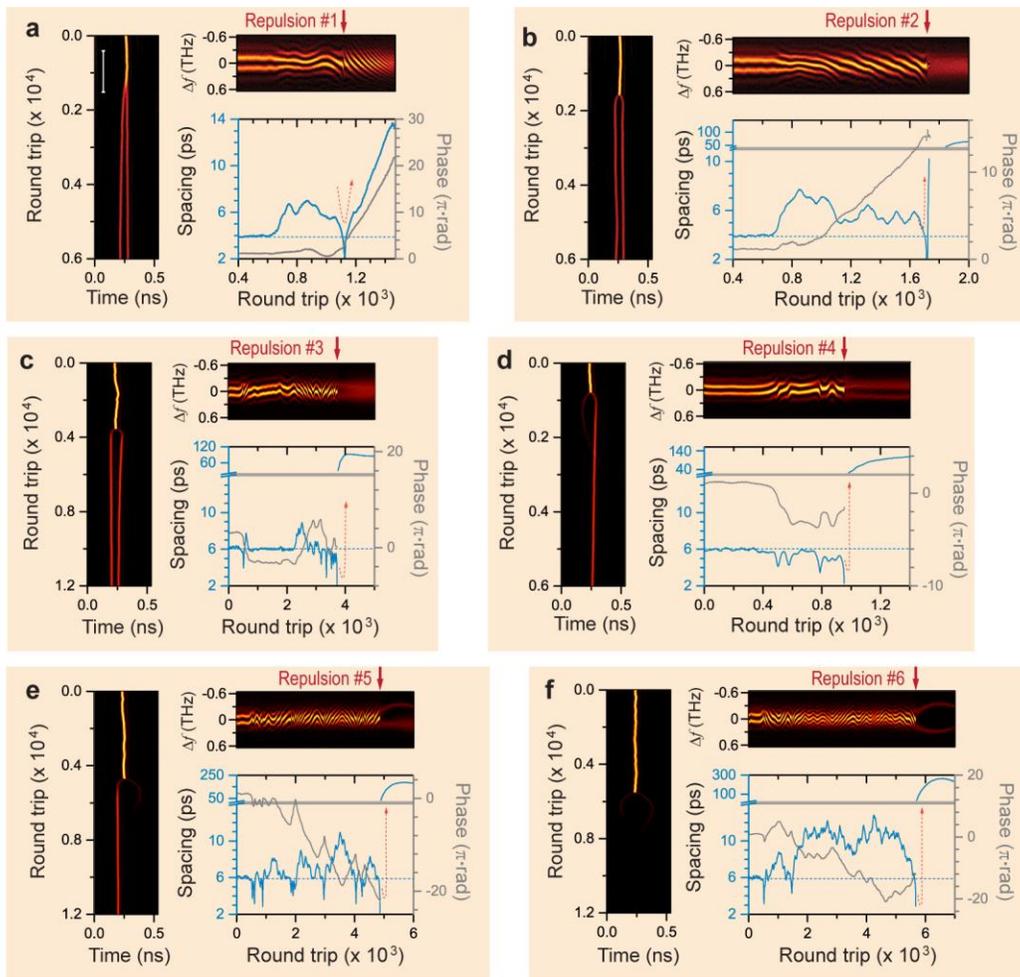

**Supplementary Figure 13** | **Radical repulsions during soliton-pair molecule dissociation**. Six time-slot examples **a** – **f** are illustrated, each enclosed within a yellow box. Each example include the time-domain recoding of soliton dynamics over consecutive round-trips within the time-slot (left panel), the DFT signal (right-top panel) and the retrieved spacing/phase-relation (right-bottom panel) over the selected region. **a** – **c** Soliton-pair molecule dissociation terminated by strong repulsions in the end, with different diverging velocity after the repulsion. **d**, **e** Soliton-pair molecule dissociation, with one of the soliton extinct after the radical repulsion in the end of the dissociation. **f** Soliton-pair molecule dissociation that ends up with both solitons extinct after the radical repulsion.

In some cases the radical repulsion would simply end up with one (see **d** and **e**) or both solitons (see **f**) quickly disappear. We note that the solitons did not disappear exactly during the repulsion. Instead, they were gradually attenuated afterward during the diverging motion. Therefore, we suspect that the carrier-frequency shift might influence the gain/loss balance of the soliton (due to limited gain bandwidth), and the soliton might experience net loss during the diverging motion and failed to recover before the long-range forces shifted the frequency back.

In the examples above, we note that some repulsions are so radical that the recorded DFT signal becomes difficult to analyse. In our experimental results, no spacing or phase information can be retrieved after the soliton spacing suddenly narrowed down below 2 ps since the DFT signal becomes almost flat, reaching the background noise level (see examples **c** – **f**). We plot the expanded view of the DFT signal over the repulsion regions in all the six examples (marked by the red arrow as Repulsions #1 – #6) in Supplementary Figure 14a. We can notice that the signal evolution is rather smooth in Repulsion #1 (a "mild" repulsion), while being less and less smooth from Repulsion #2 to Repulsion #6, indicating more and more radical repulsions. Actually, we noticed that these radical repulsions could complete within only a few round-trip, indicating that the dramatic changes can occur even within one round-trip, which is beyond the analysing capability of the conventional DFT method. We plot for example the round-trip by round-trip DFT-signal for Repulsion #6 in Supplementary Figure 14b. The DFT signal before round-trip Nr.75 still gives clear interferometric fringe (with a central fringe peak that indicates an in-phase-relation), while in Nr.76 the signal becomes almost flat, from which we can hardly know the exact motion of the two solitons. The solitons themselves might experience dramatic distortions[9] under such radical repulsion at intimate spacing, and could no longer be regarded as particle-like wave-packet that are still maintaining their profile.

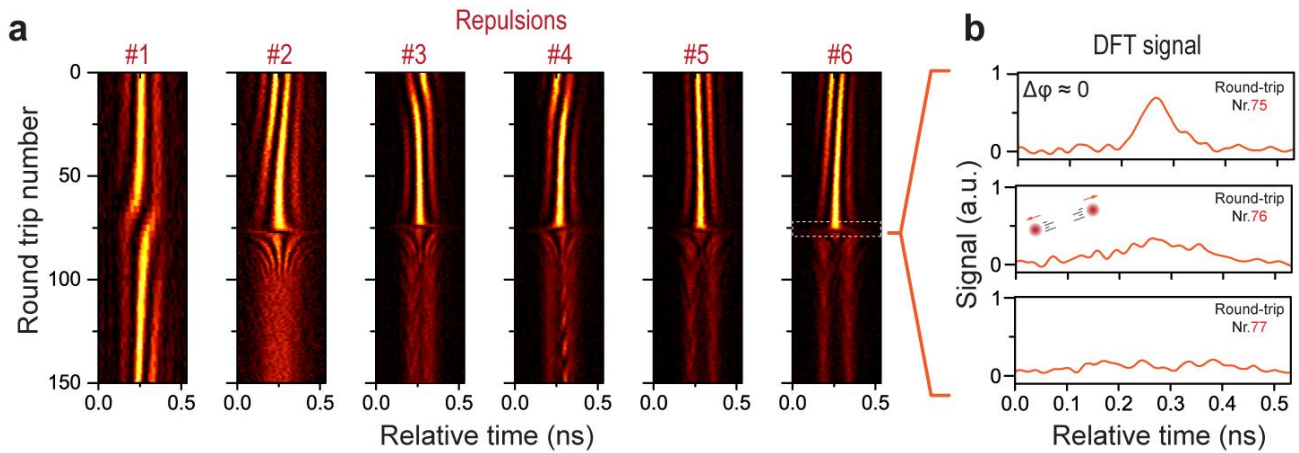

**Supplementary Figure** 14 | **a** Expanded views of the DFT signal at regions where soliton repulsions occurred during the dissociation marked in Supplementary Figure 13 as Repulsion #1 – #6. In each case, we plot the DFT signal over 150 consecutive round trips where the repulsion occurred in the middle of selected time-span. **b** DFT signal over 3 consecutive round trips at which radical repulsion occurred in the example of Repulsion #6.

According to conventional theory, the two solitons tends to attract when being in phase while repulsing when being out of phase. We can notice from all the DFT signal from Supplementary Figure 14a that the two soliton always tend to become in-phase (with a major interferometric peak in the middle of the DFT signal) while getting attracted towards each other, in accordance to the conventional theory. However, such attraction force immediately convert into radical repulsion avoiding the two soliton to merge into one. The repulsion of two optical solitons in fibres at intimate spacing has been noticed since decades[10], while the exact mechanism behind this phenomenon remains unclear. Soliton self-frequency shift[11] due to Raman effects has been suspected to cause such shift of phase-relation. The two solitons with in-phase-relation tend to attract each other, leading to increased amplitude out of the constructive interference, which then caused stronger Raman shift. A mutual shift of the soliton carrier-frequency $\Delta v$ of the soliton-pair with spacing of $\Delta t$ would lead to a relative-phase shift[10] of $\Delta\varphi = 2\pi \Delta v \Delta t$. As a consequence, the in-phase-relation that cause the soliton attraction would be quickly shifted to out-of-phase-relation, turning the internal forces into radical repulsion. In our case, we suspect that the peak-power clamping effect[12,13] provided by the NPR-effect, as a critical nonlinear dissipative process of the mode-locked laser cavity, may also contribute to preventing the two soliton merge into one with enhanced peak power. The repulsions during the soliton-triplet molecule synthesis shown in panel (v) – (viii) of Supplementary Figure 10 may also result from similar mechanism as the two-soliton ones, thought with more complicated trajectories. Currently we could not track the exact evolution of the multi-soliton dynamics within one round-trip, more theoretical and simulations would be needed to completely understand the repulsion mechanism of multi-soliton interaction.

*Metastable molecular states.* Within the stochastic trajectories of the dissociation (partially shown in Figs.4 and 5 in the article), we notice that the two solitons can transiently reside at some stable spacing and phase-relation other than the original molecular state[14,15]. We called these states as metastable states, which might be attractors with less stability in the current nonlinear system, compared with the original soliton-pair molecule. In Fig.5e of the main article, we can already notice a metastable state at spacing of ~12 ps with fixed phase of ~$\pi$ during the dissociation of the soliton molecule with original spacing of 4.9 ps. In Supplementary Figure 15, we present four examples over the parallel reactors within the same dissociation of soliton-pair molecules. The time-domain recording over each of the time-slots are given in **a** (panel (i) – (iv)) while the corresponding DFT-signal and the retrieved spacing and phase-relation are given in **b** – **e**. The metastable states within the dissociation trajectories of each example are marked by grey arrows. Some metastable states are found to be shared by all the reactors during the dissociations. As a brief illustration, the examples given in **b** and **c** are share the metastable spacing at ~6.2 ps, while examples given in **d** and **e** share the metastable state at ~5 ps (the metastable spacing are marked by blue dashed lines).

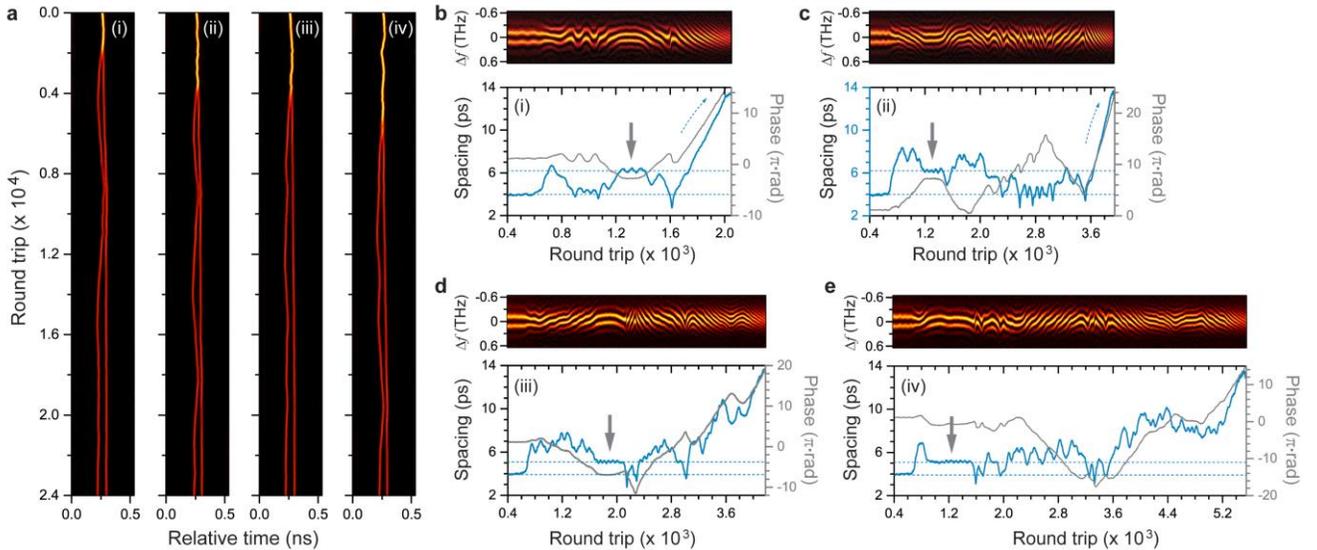

**Supplementary Figure 15 | Soliton-pair molecule dissociation with transient metastable states. a** The 4 panels (i) – (iv) are time-domain recordings of soliton-pair molecule dissociation in 4 selected time-slots over 24000 round trips. The pump power perturbation is introduced at round trip of ~700. **b** – **e** The corresponding DFT signal and retrieved spacings and phase-relations for examples in panel (i) – (iv) in **a**. The region where the two soliton reside transiently one some metastable states are marked by a grey arrow in each plot. The blue dashed lines in **b** – **e** mark the original molecular spacing and the metastable spacings.

*Random-walk-like motion.* The stochastic motion of the solitons in during the molecule dissociation are found to resemble one-dimensional random walk with fixed length of motion step (See the example of Fig.5e in the main article). Here we provide a few more examples in Supplementary Figure 16 which exhibit random-walk-like trajectories that feature randomly oscillating spacing with step-wise change during the dissociation. The minimum step length (spacing change) is measured to be ~0.6 ps, and one such step takes ~50 round trips. For comparison, we numerically generated one-dimensional random walk trajectories and plot one example in Supplementary Figure 16d. Each step of walk has a length of 1 and a random direction (+1 or −1). The deviation gradually increases after hundreds of steps, highly resembling the trajectory of the soliton motion retrieved from DFT signal in Supplementary Figure 16a – c.

Such random-walk-like behavior is probably due to the phase-dependent interactions between the solitons. The dispersive waves shed from the solitons, which is enhanced after the induced perturbation and has different phase accumulation rate from the soliton, might also be responsible for such stochastic soliton interaction. Once the soliton spacing changes due to their interactions, their relative phase changes accordingly, so as the direction of their interaction force, leading to random and oscillatory motion. The exact mechanism that drives the random-walk-like motion remains an open question. In addition, this random-walk-like motion is mostly accompanied with the radical repulsion at the boundary of molecular-bond spacing, metastable states, and eventually the long-range trapping potential, making the exact motion highly complicated. We expect that DFT analysis with higher resolution would be needed in order to retrieve finer trajectory of the soliton motion for further investigations, and the advantage of such parallel reactor scheme can then be further enhanced.

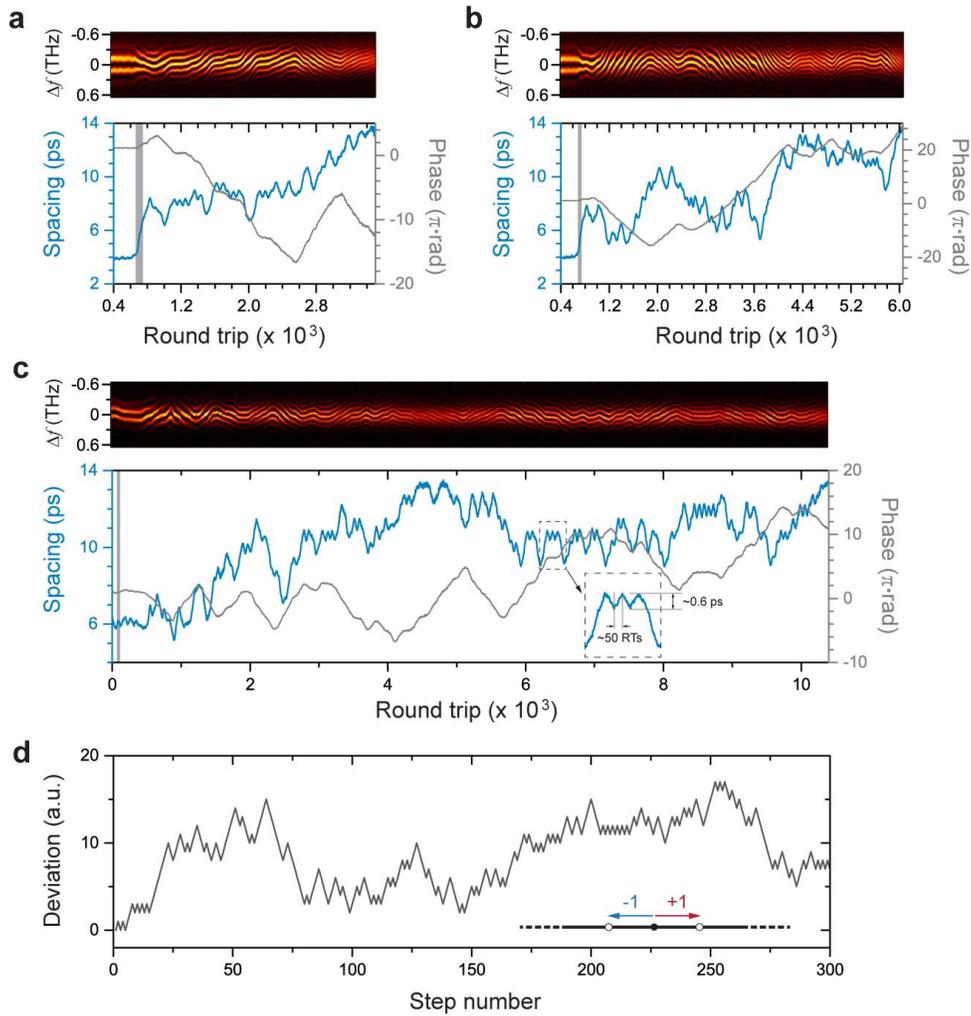

**Supplementary Figure 16 | Random-walk-like motion during soliton-pair molecule dissociation**. **a** – **c** DFT signal evolution of a soliton-pair molecule dissociation within one time-slot, The grey line indicate the induced perturbation. The initial soliton spacing is 5 ps in **a**, **b** and 6 ps in **c**. **d** A numerically generated 1D random walk over 300 steps, with fixed step length of 1 and random direction (+1 or -1). The randomly oscillatory deviation is plotted versus the step number.

**More examples of soliton-triplet molecule dissociation**

The examples shown in Panel (iv) and (v) of Fig.5a give two possibilities for dissociation of the soliton-triplet molecules within the parallel reactors. In our experiment, the soliton triplet mostly dissociated into a soliton-pair and a single-soliton, which are bound at long-range in the end of the process as sketched in Supplementary Figure 17a. The dissociation of soliton triplets are also highly stochastic, as partially shown in Supplementary Figure 17b and c, with time-domain sequence and the corresponding DFT signal over 8 consecutive time-slots over the initial ~29000 round trips. For comparison, a few time-slots are initially prepared to host only one soliton-pair molecule or one single-soliton. The global perturbation is introduced as a ~1-dB decrease of the cavity loss in at round-trip number of ~400. We can also readily notice the group-velocity discrepancies between all these different solitonic elements during the dynamic process, which are eventually synchronized by the parallel reactors.

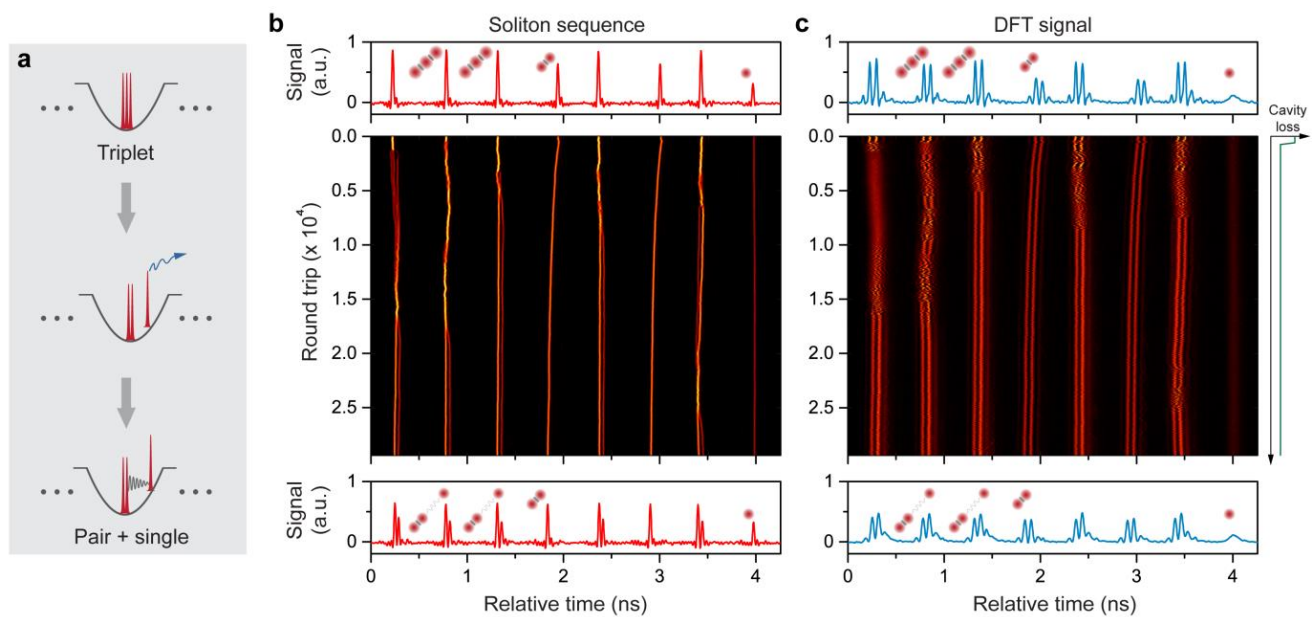

**Supplementary Figure 17 | Dissociation of soliton-triplet molecules**. **a** Schematic of a typical dissociation of an soliton-triplet molecule within the parallel reactor. One of the molecular bond break down, releasing a single-soliton which drifts away from the remaining soliton-pair and eventually forms a long-range binding with the soliton-pair. **b** The time-domain recording of 8 consecutive time-slots (out of 160) over ~29000 round trips in which dissociation of soliton triplets occurred. Note that there are a few time-slots containing either a soliton-pair or a single-soliton, in order to show the group-velocity discrepancies between different solitonic elements during the dissociation. **c** The corresponding DFT signal of the time-domain recording in **b**.

Even for the simplest case, in which the soliton triplet dissociate into a soliton-pair and a single-soliton, two possibilities exist depending on which of the two molecular bonds within the soliton triplet breaks down during the dissociation. We can easily find both possibilities in the parallel reactors under a single dissociation reaction. Two exemplary trajectories retrieved from DFT signal are plotted in Supplementary Figure 18. In both cases, one of the side soliton breaks away from the remaining soliton-pair which still maintained their original molecular bond. For similar reasons as in the synthesis, here we could only retrieve some rather simple and smooth trajectories. Further investigations into the mostly complicated trajectories of soliton-triplets would require DFT signal with higher resolution and precision, and we expect that all the four primary features summarized above for soliton-pair molecule would also apply for soliton-triplet molecules.

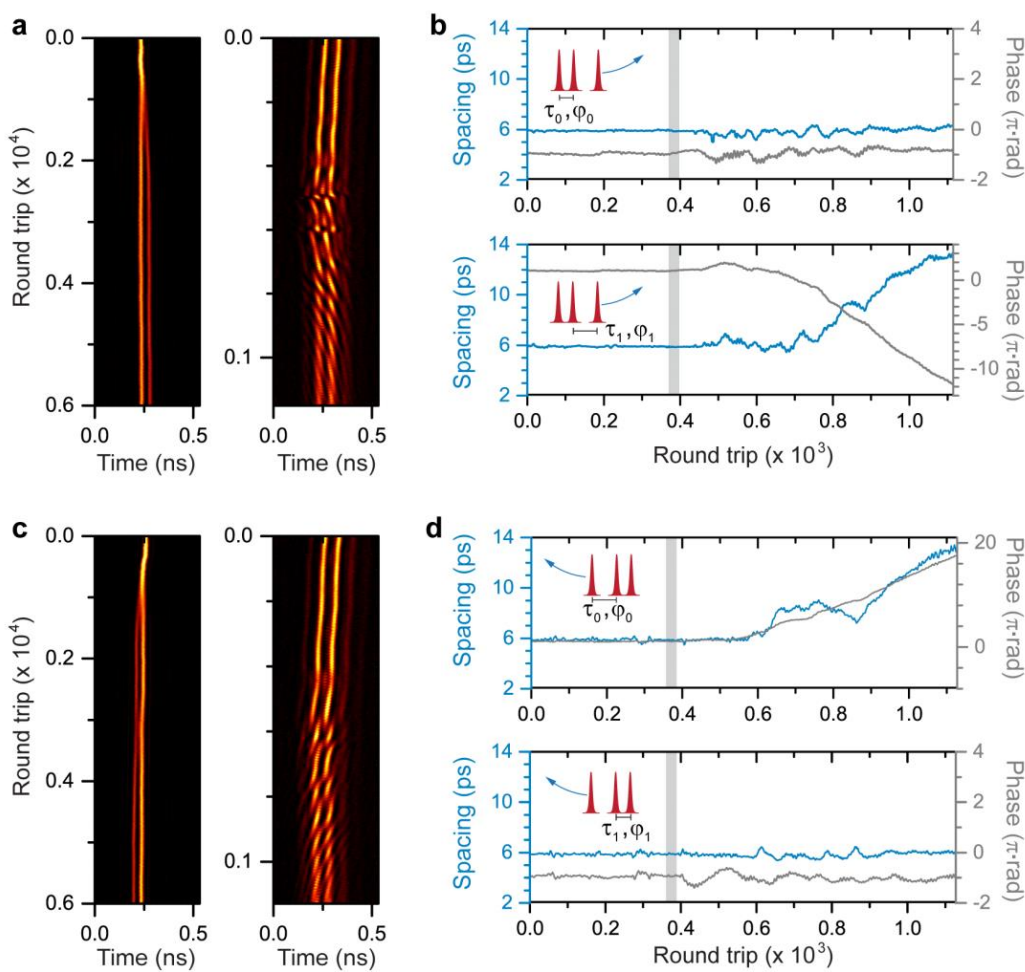

**Supplementary Figure 18 | Two typical dissociation trajectories of soliton-triplet molecules. a** The time-domain soliton sequence (left) and the corresponding DFT signal (right) of the selected time-slots over the initial 6000 round trips. **b** The retrieved spacings and phase-relations between the three solitons during the dissociation. The single-soliton on one side of the triplet pattern breaks away from the remaining soliton-pair which keeps their spacing and phase-relation. **c** and **d** illustrate the other possibility, in which the single-soliton on the other side of the soliton triplet break away from the remaining the soliton-pair.

Apart from the smooth dissociation shown above, some radical break-down of the soliton-triplet molecules are also observed within some other parallel reactors during the same dissociation. In these cases, both molecular bonds break down, which can occasionally lead to extinctions of the constituting solitons, similar to the cases shown in Supplementary Figures 10 and 13. Three typical trajectories with DFT signal are shown in Supplementary Figure 19, which present a few possibilities of such radical dissociation (with 3, 2 or only 1 soliton left over after the dissociation). We can notice from the DFT signal that the dissociation is completed by a radical repulsion between the solitons, similar to the cases shown in Supplementary Figure 14a. The three-soliton repulsion is currently difficult to analyse due to limited resolution of the DFT method and non-unique fitting parameters that result in undetermined spacing and phase-relation between all three solitons during such radical reaction.

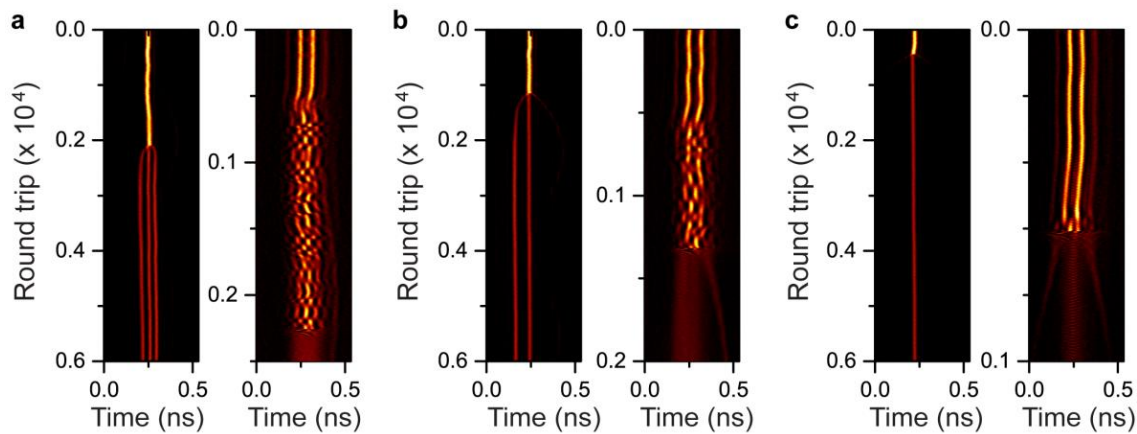

**Supplementary Figure 19 | Dissociation of soliton-triplet molecules with radical repulsion. a** Dissociation of a soliton triplet that ends up with repulsions between all three solitons. The left panel show the time-domain soliton sequence evolution over the initial 6000 round trips and the right panel shows the corresponding DFT signal. **b** Dissociation of a soliton triplet that end up with losing one of the soliton after an radical repulsion between the threes solitons. **c** Dissociation of a soliton triplet that ends up with losing two solitons after an radical repulsion between the three solitons. These time-slots are from the same experiment result as in Fig.5a of main article and Supplementary Figure 17 and 18 above.

## Supplementary Note 6: Individual control technique

### Generation of external addressing pulses

The individual control over the soliton reactions in selected reactors is realized by launching external pulses which are generated by modulating a single wavelength laser (as shown in the dashed red box in Supplementary Figure 1). The single wavelength laser emits CW laser at 1550 nm with a linewidth of 2 kHz and an average power of ~30 mW. Then the CW light is modulated by an electro-optic modulator (EOM) with a voltage bias and pulsed voltage signal generated by a programmable pulse generator. The pulse pattern we programmed has 256 slots with a clock rate that matches the repetition rate of the mode-locked laser that is also operating at exactly 256-th harmonic order. We fill 10 to 20 slots in one pattern period with electrical pulses in order to overlap with selected time-slots in the laser cavity. The electrical pulse duration (and therefore the optical pulse duration) is set to be 200 ps, which would be long enough to cover the long-range bound solitons in the time-slots of the mode-locked laser while with sufficient amplitude to introduce cross phase modulation. The optical pulses generated by the EOM are then amplified after two stages (with a 1-nm band-pass filter in between to suppress ASE) to reach an average power of 2 W. Then each pulse has ~10 nJ energy, with peak power of 50 W. Then the launching of these pulses are controlled by using an optical switch controlled by the function generator, which has a response time of 300 ns, an extinction ratio of 20 dB, and an intrinsic loss of ~4 dB. A square voltage pulse is applied to the optical switch so as to launch a finite number of repeated external pulse pattern into the laser cavity. After a second band-pass filter (1 nm) and an isolator, the transmitted pulses are then adjusted in polarisation states by the FPC/Polariser/FPC combination before they are launched into the cavity through the 50/50 output coupler. The external pulses should interact with the intro-cavity pulses only through XPM during their direct overlapping, without perturbing the laser gain section. Therefore, we first adjusted the polarisation state of the external pulses such that they can be eliminated by the intra-cavity polariser after sufficient co-propagation with the solitons. Importantly we inserted additionally ~6-m-long SMF after the 50/50 coupler in order to enhance the interaction by increasing the co-propagation distance for the external and internal pulses. The peak power of the external pulse that is launched into the cavity (after the 50/50 coupler) is attenuated to ~10 W, which is still comparable to that of the intra-cavity soliton (~25 W, if there are the two solitons in each time-slots of 1.88 GHz repetition rate, giving an average power of 100 mW at 50/50 output coupler). Therefore the perturbation remains significant during their direct overlapping.

In order to obtain clean DFT signal of the interacting solitons under the addressing pulses modulation, we inserted a second output coupler (90/10) just before the 50/50 coupler specifically for DFT signal detection. Since the addressing pulses are eliminated by the intra-cavity polariser, only the soliton signal would reach this second coupler. As a result the low-amplitude DFT signal during the soliton interactions due to the individual control will be free from any smearing caused by the addressing pulses.

**Mechanism of individual control**

The addressing pulses that directly overlap with solitons in specific time-slots can significantly perturb the existing balance of forces between the solitons, leading to desired soliton reactions, even though the laser parameters remains unvaried during this process. For synthesis of soliton molecule, the addressing pulse pattern is set to be synchronized with intra-cavity soliton in terms of the cavity round-trip frequency, so that each addressing pulse can constantly overlap with the target time-slot without relative drifting. Such perturbation would cause the two long-range bound solitons to collide. In the examples shown Fig.6b of the main article, such collision end up with successful formation of soliton-pair molecule. In most other time-slots of the parallel reactors during the same experiment, however, the collision failed to establish the molecular bond, and the two solitons repel each other again into the long-range bound states, as shown in Supplementary Figure 20a. This is a reasonable result considering the multiple collisions required for soliton molecule synthesis under global control (e.g. see Fig.2 of the main article), since the initial phase-relations between the interacting solitons vary from time-slot to time-slot.

The mechanism of such induced collision by external addressing pulse could possibly be understood in the following way. The two solitons were in stable balance that the repulsion and attraction of one soliton exerted upon the other create a secondary potential for a second soliton, which has been explained in detail in our previous work (see Ref[8] and Supplementary Figure 20b). When the addressing pulses are launched into the cavity and overlap with this target time-slot, the two solitons, due to their large spacing, would experience different cross-phase modulation. The soliton (at $t_2$) that rides on the peak of the addressing pulse would be delayed due to the XPM-induced nonlinear refractive index compared with the other soliton (at $t_1$) riding at the tail of it. Therefore, the soliton at $t_2$ would have relative motion towards the soliton at $t_1$, appearing to experience addition attraction between them other than the one caused by the optoacoustic effects. Meanwhile, the repulsive force between the solitons due to dispersive wave perturbation could also be perturbed by the overlapping of the addressing pulses. Consequently, the effective attraction between the two solitons exceeds their repulsion and the two solitons start to move towards each other (see Supplementary Figure 20c). When the two solitons become close enough, they began to follow the stochastic trajectories due to phase-sensitive interactions, similar to the reactions under global-control approaches.

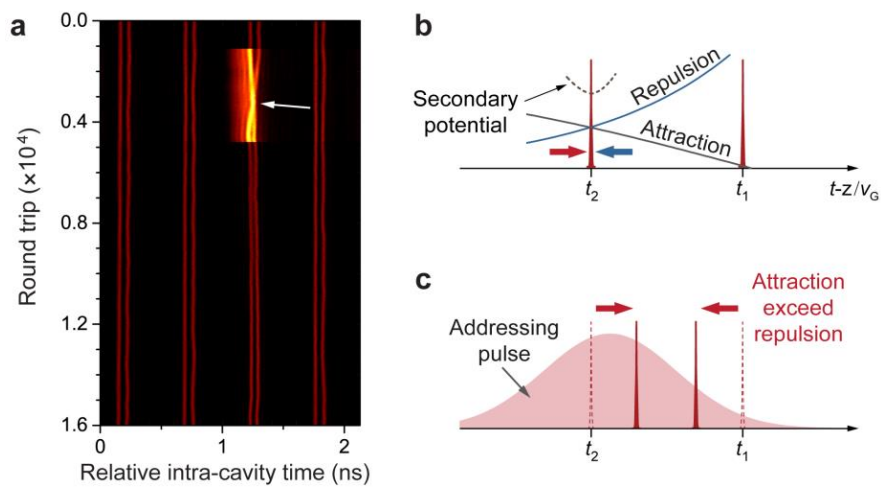

**Supplementary Figure 20 | Induced soliton collision by external addressing pulses**. **a** Time-domain recording of a few time-slots during the individual control experiment (same as the one shown in Fig.6b of the main article). These time-slots all host long-range double solitons, while one of them experience perturbations from the external addressing pulses that constantly overlap with the two solitons in it, leading to soliton collisions. **b** Schematic of the balanced forces between the long-range bound solitons before the external perturbation. **c** The overlapping of the addressing pulses with the two solitons causes additional attraction that exceeds the repulsions between them, leading to subsequent soliton collisions.

For dissociation of soliton molecules in selected time-slots, we found in experiments that the most efficient strategy is to launch the addressing pulse with a slight asynchronized repetition rate (in our case ~20 Hz), so that the addressing pulses would traverse across the target soliton-pair molecules instead of constantly overlapping them (see the experimental recording in Supplementary Figure 21a). We can infer from the DFT signal (Supplementary Figure 21b) that the soliton molecule was actually first squeezed into a much narrower spacing (~2 ps) well below the molecular spacing (~7 ps), which then triggered a radical repulsion between the two solitons, similar to the cases shown in Supplementary Figure 14. Then the dissociation was quickly completed after the radical repulsion and long-range binding was established.

The way that the traversing external pulses perturb the individual solitons within the molecule probably resembles the way that the dispersive wave perturb the solitons that exert long-range repulsive forces[8,16], although with a much stronger strength. Through the unbalanced XPM effect, the traversing pulses pushes one soliton towards the other and cause the radical repulsion afterwards. Another way to view this process is to analyse the relative phase of the two solitons during the dissociation. The relative phase under the balanced state of the molecular bond is ~π. The external addressing pulses, when traversing over the soliton-pair, gradually change the relative phase towards to 2π (or 0), which turned the forces between the solitons into net attraction[10] and lead to the sudden narrowing of soliton spacing (see Supplementary Figure 21d). Then the radical repulsion emerged, leading to collapse of the molecular bond and diverging of the two solitons away from each other.

The significance of this traversing-pulse strategy is that it has turned a probabilistic event − the radical soliton repulsion observed only in the stochastic trajectories of soliton reaction (see Supplementary Figure 13) − into a deterministic event that can be triggered on demand to quickly achieved the desired dissociation. Currently, the optimized frequency-offset between the addressing pulse and the intra-cavity soliton is determined empirically. We expect to develop a theoretical model to fully interpret the perturbation mechanism, which could also help us to further improve the individual-control method.

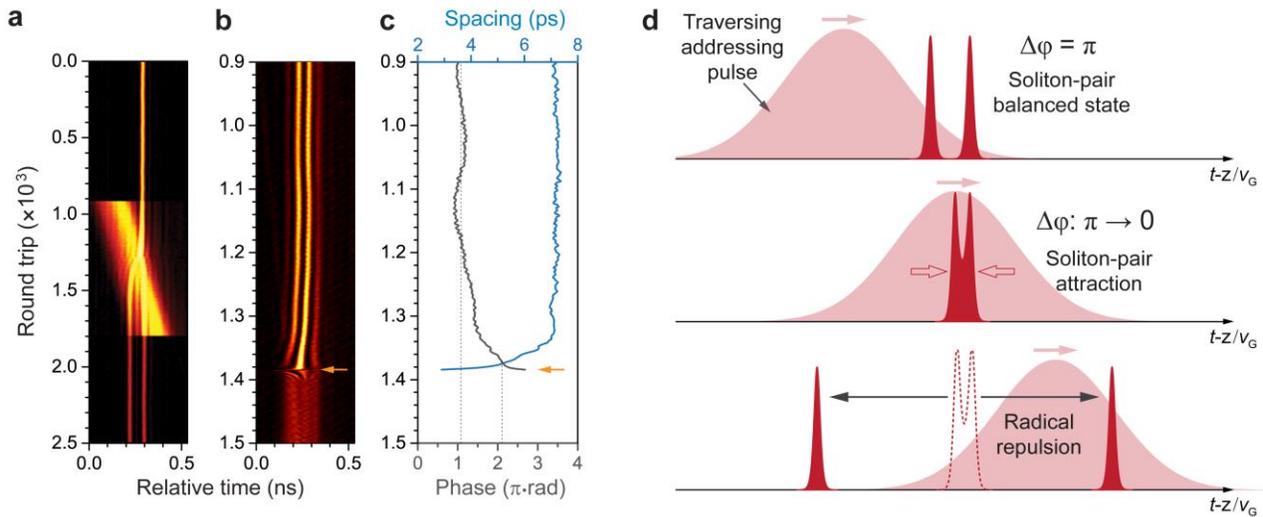

**Supplementary Figure 21 | Soliton-pair molecule dissociation using traversing external pulses. a** Time-domain recording of the soliton-pair molecule dissociation triggered by the external pulse with slightly lower repetition rate. **b** DFT signal and **c** the retrieved spacing and phase-relation between the two solitons during the dissociation. A radical repulsion occurred (marked by the yellow arrow) before the two solitons completely dissociate. **d** Schematic of the possible mechanism of the dissociation through addressing-pulse perturbation. The addressing pulses traverse through the target soliton-pair, cause a flipping of the phase-relation (from out-of-phase to in-phase) followed by a sudden narrowing of the soliton spacing. Then a radical repulsion occurred which quickly push the two solitons away from each other.

## Supplementary Movie 1: Soliton molecule synthesis in parallel reactors

Cylindrical-coordinate plot of the parallel reactors (with 195 time-slots) during the synthesis of the soliton-pair molecules. Left panel is the time-domain recording and the right panel is the corresponding DFT signal for each frame of recording. The frame rate is 5 kHz, corresponding to ~1900 round trips per frame. The synthesis was initiated by a slight increase (~8%) of the pump power.

## Supplementary Movie 2: Soliton molecule dissociation in parallel reactors

Cylindrical-coordinate plot of the parallel reactors (with 195 time-slots) during the dissociation of the soliton-pair molecule. Left panel is the time-domain recording and the right panel is the corresponding DFT signal for each frame of recording. The frame rate is 20 kHz, corresponding to ~480 round trips per frame. The dissociation was initiated by a slight decrease (~8%) of the pump power. Note that a few soliton molecules lost one of the constituting soliton after the dissociation, due to similar phenomena illustrated in Supplementary Figure 13c and d.

## Supplementary References